\documentclass[manuscript]{aastex}

\slugcomment{accepted by ApJ, 20 May 2011}

\shorttitle{Ionisation  in atmospheres of very low-mass objects}
\shortauthors{Helling et al.}

\begin{document}


\title{Ionisation in atmospheres of Brown Dwarfs and extrasolar planets
       II Dust-induced  collisional ionization}


\author{Ch. Helling\altaffilmark{1} and M. Jardine\altaffilmark{1}}
\affil{ 1: SUPA, School of Physics \& Astronomy, University of St Andrews, North Haugh, St Andrews,  KY16 9SS, UK}
\email{ch80@st-andrews.ac.uk}

\author{F. Mokler\altaffilmark{2,3}}
\affil{2: Max-Planck-Institut f\"ur extraterrestrische Physik (MPE), Gie{\ss}enbachstr. 1, D-85748 Garching}
\affil{3: Max-Planck-Institut f\"ur Gravitationsphysik, Callinstra{\ss}e 38, 30167
Hannover, Germany}




\begin{abstract}
Observations have shown that continuous radio emission and also
sporadic H$\alpha$ and X-ray emission are prominent in singular, low-mass
objects later than spectral class M. These activity signatures are
interpreted as being caused by coupling of an ionised atmosphere to
the stellar magnetic field. What remains a puzzle, however, is the
mechanism by which such a cool atmosphere can produce the necessary
level of ionisation. At these low temperatures, thermal gas processes
are insufficient, but the formation of clouds sets in. Cloud
particles can act as seeds for electron avalanches in streamers that
ionise the ambient gas, and can lead to lightning and indirectly to magnetic field coupling, a combination of processes also expected for
protoplanetary disks. However, the precondition is that the cloud particles are
charged.

We use results from {\sc Drift-Phoenix} model atmospheres to
investigate collisional processes that can lead to the ionisation of
dust grains inside clouds.
We show that ionisation by turbulence-induced dust-dust collisions is the most
efficient kinetic process. The efficiency is highest in the inner
cloud where particles grow quickly and, hence,  the dust-to-gas ratio is high.
Dust-dust collisions alone are not sufficient to improve the magnetic
coupling of the atmosphere inside the cloud layers, but the charges supplied
either on grains or within the gas phase as separated
electrons can trigger secondary non-linear processes. Cosmic rays are
likely to increase the global level of ionisation, but their
influence decreases if a strong, large scale magnetic field is
present as on Brown Dwarfs.

We suggest that although thermal gas ionisation  declines in
objects across the fully-convective boundary, dust charging by
collisional processes can play an important role in the lowest mass objects. The
onset of atmospheric dust may therefore correlate with the anomalous
X-ray and radio emission in atmospheres that are cool, but charged
 more than expected by pure thermal ionisation.
\end{abstract}


\keywords{Brown Dwarfs, atmospheres, dust, ionisation, magnetic coupling}



\section{Introduction}

Both Brown Dwarfs and planets form clouds in their atmospheres. The
radial extent of these clouds is mainly a function of gravity (Helling
et al. 2010). High altitude clouds in the giant gas planet HD~189733b
were inferred from transit measurements (Pont et al. 2008, Sing et
al.2009). HD~189733b was the first exo-planet to be detected in
polarized light which is likely to be caused by Rayleigh scattering on
sub$\mu$m grains (Berdyugina et al. 2011). Lecavelier Des Etangs et
al. (2008), too, suggest sub$\mu$m dust grains as source of Rayleigh
scattering in their analysis of the HD~189733b transit spectrum. 
The understanding that clouds are present in Brown Dwarf atmospheres  was triggered by the disappearance of metal-oxide band (like TiO and VO) with decreasing effective temperature (see Kirkpatrick 2005). The comparison with
model atmospheres showed that cloudless models completely fail to explain the NIR colors and L-dwarf spectra (Chabrier et al. 2000, Burrows et al. 2006, Saumond \& Marley 2008). By today,   various approximations of clouds as
opacity source are in application (e.g. Burgasser at al. 2010, Currie et
al. 2011) as the existence of clouds in Brown Dwarfs is well accepted. Photometric variability in L- and in T-dwarfs has been
attributed to cloud evolution (e.g. Bailer-Jones 2008, Artigau et
al. 2009), a finding which is supported by our work on turbulent dust
formation (Helling et al. 2001, 2004).  Goldman et al. (2008),
however, could not confirm spectroscopically the so-called
'cloud-braking' across the L-T transition based on their model
atmosphere interpolation.  


Scholz \& Eisl\"offel (2004) argue that very low-mass objects
($0.08\,\ldots\,0.25$M$_{\odot}$) show co-rotating, magnetically
induced spots, which would indicate the presence of localised magnetic
field concentrations.  Objects at the boundary between M-dwarfs and
Brown Dwarfs ($0.08\,\ldots\,0.09$M$_{\odot}$) were studied by
several groups considering the activity change across the substellar
border. The presence of X-ray  and H$\alpha$ emission (Berger et al.
2008, Reiners \& Basri 2008) argues for a magnetically-heated corona,
but the associated increase in radio emission with decreasing mass
suggests that the processes that are energising such a corona differ
from those on higher-mass stellar objects (Hallinan et al. 2006, 2008;
Antonova et al. 2008). In this case, the solar paradigm for coronal heating, which involves
the build-up and release of magnetic stresses in the corona, may be
ineffective.

We have argued in Helling et al. (2010) that in the atmospheres of
Brown Dwarfs and exoplanets, rapid ionisation events can occur, even
in a cloud that is globally neutral and obeys dust-gas charge
equilibrium. This can only happen, however, if the grains within the
dust cloud are ionized. As one possible process, electron avalanche
processes in the electric field of charged grains than lead to the
formation of a streamer resulting in $10^{13}\ldots 10^{14}$ cm$^{-3}$
free charges per initial electron (Dowds et al 2003, Li et
al. 2007). Streamers are a growing ionisation front that propagates
into non-ionised matter and are present in lightning ladders and sprites.

The timescale on which these charges could recombine (the Coulomb
recombination time scale) is longer than the time to establish a
streamer over a large fraction of the cloud (Helling et
al. 2010). Hence, a considerable degree of ionisation builds up in
this time window during which the electrons could couple to the
large-scale magnetic field.
 
 In this paper we evaluate the underlying assumption in this scenario
 - that the grains in substellar clouds are ionised.  We concentrate
 on the influence of dust dynamics (sedimentation, turbulence) and
 kinematics (collisions) on the ionisation state of Brown Dwarf and
 giant planet atmospheres.  Section~\ref{sec:cf} briefly summarises
 our kinetic dust cloud model and the results relevant for the topic
 of this paper.  In Secs.~\ref{ss:wf},~\ref{s:colion} we study the
 collisional energies of dust-gas and dust-dust collisions due to dust
 dynamics, and compare these results to the ionisation potential of
 solid compounds for an example of a brown dwarf and a gas planet
 atmosphere. Section~\ref{ss:regimes} discusses potential lightning regimes in these atmospheres and
Sect.~\ref{sec:concl} presents our conclusions.

\section{Cloud formation model}\label{sec:cf}

\begin{figure}
\resizebox{14cm}{!}{\includegraphics{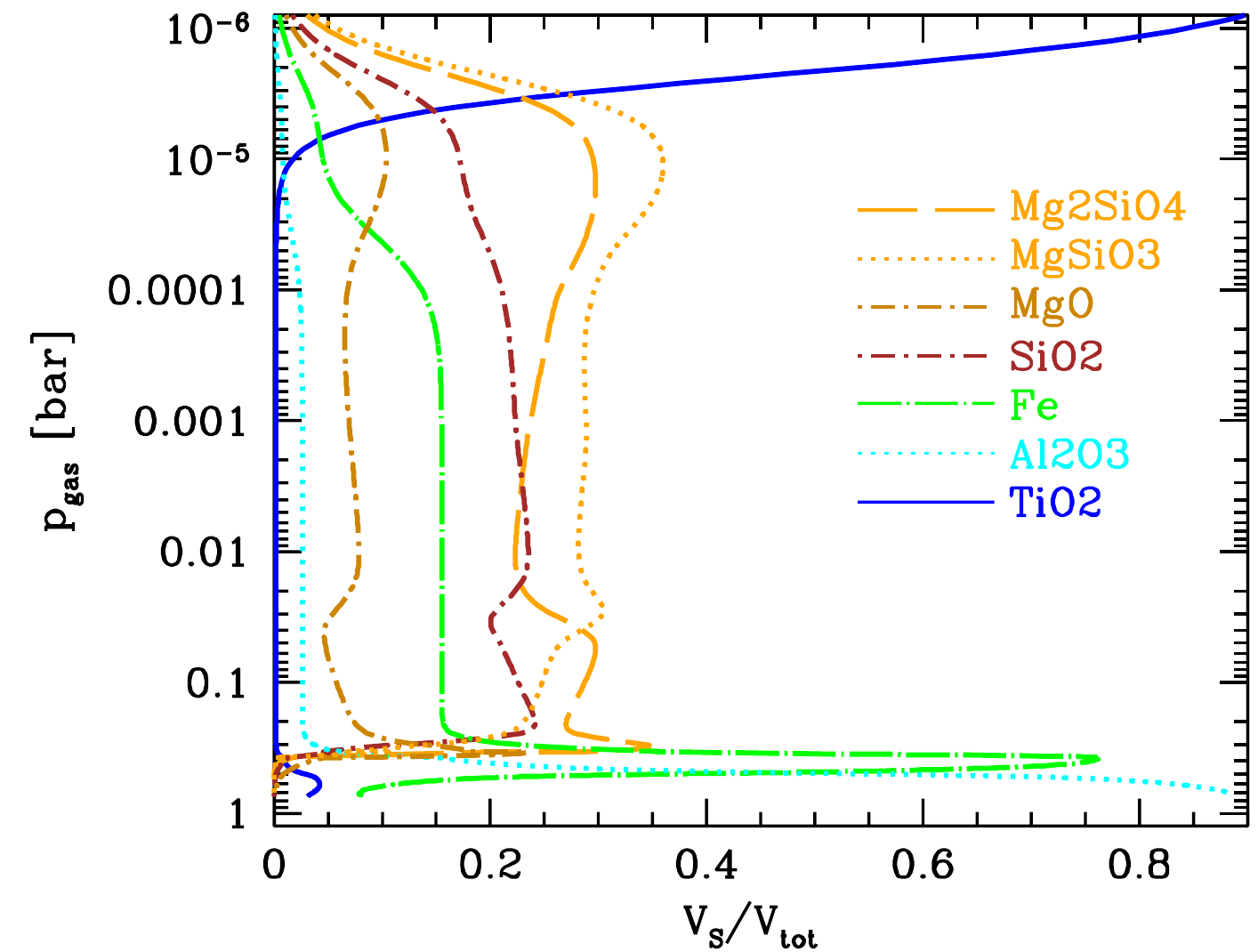}}\\
\hspace*{1.3cm}\resizebox{14.5cm}{!}{\includegraphics{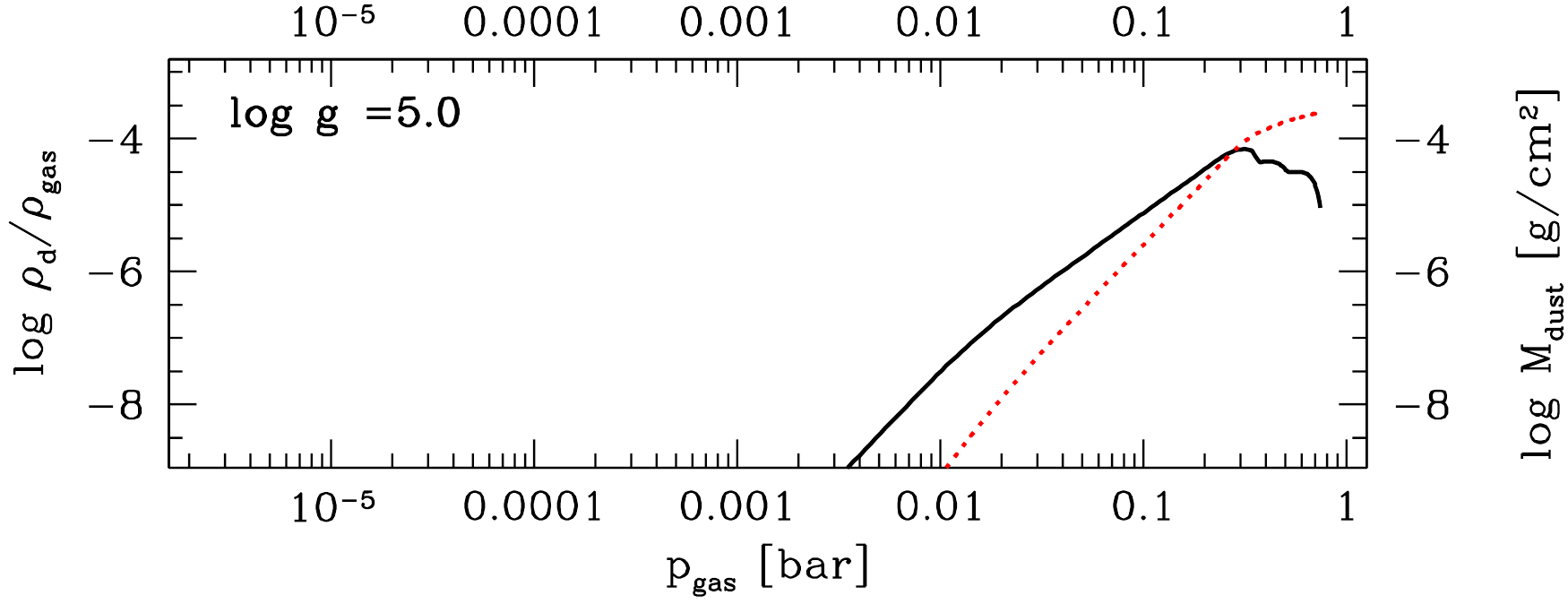}}\\
\caption{\small Material composition in volume fractions $V_{\rm
 s}/V_{\rm tot}$ ($V_{\rm s}$ - volume of a solid s, $V_{\rm tot}$ -
 total grain volume) of the dust cloud grains at a gas pressure $p_{\rm
 gas}$ in the atmosphere (top).  The bottom panel shows the
 dust-to-gas ratio $\rho_{\rm d}/\rho_{\rm gas}$ (solid line) and the
 dust mass column density $M_{\rm dust}$ [g\,cm$^{-2}$] (dotted
 line).  All quantities are results of the hydrostatic {\sc Drift-Phoenix}
 atmosphere simulation for T$_{\rm eff}=1600$K, log(g)=5.0 and solar
 metallicity. The gas pressure $p_{\rm gas}$ is therefore a measure for
 atmospheric altitude (compare e.g. top panels in Fig.~2 in Witte,
 Helling \& Hauschildt 2009). }
\label{fig:Vd5}
\end{figure}

\begin{figure}
\resizebox{14cm}{!}{\includegraphics{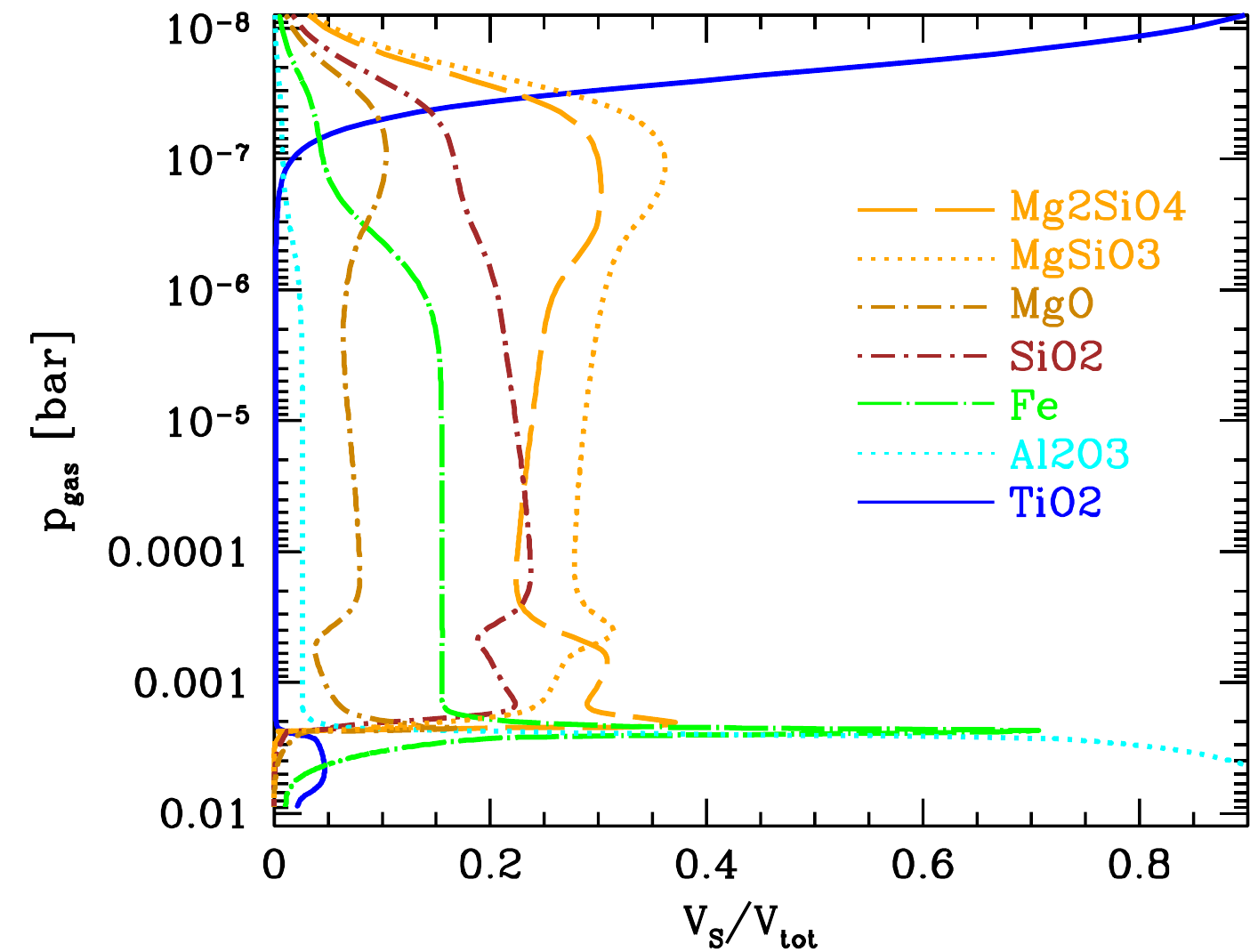}}\\
\hspace*{1.3cm}\resizebox{14.5cm}{!}{\includegraphics{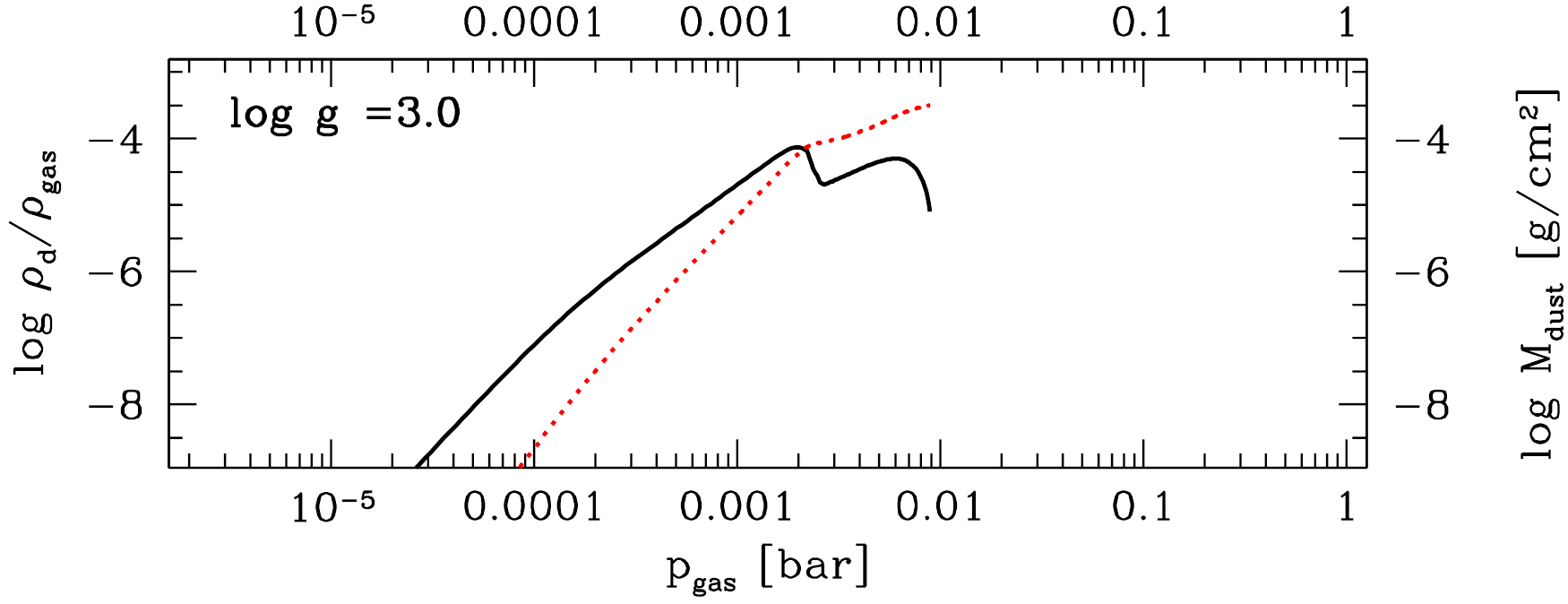}}\\
\caption{\small Same like Fig.~\ref{fig:Vd5} but for an
 atmosphere simulation for T$_{\rm eff}=1600$K, log(g)=3.0 and solar
 metallicity.}
\label{fig:Vd3}
\end{figure}

\begin{figure}
\begin{center}
\resizebox{11cm}{!}{\includegraphics{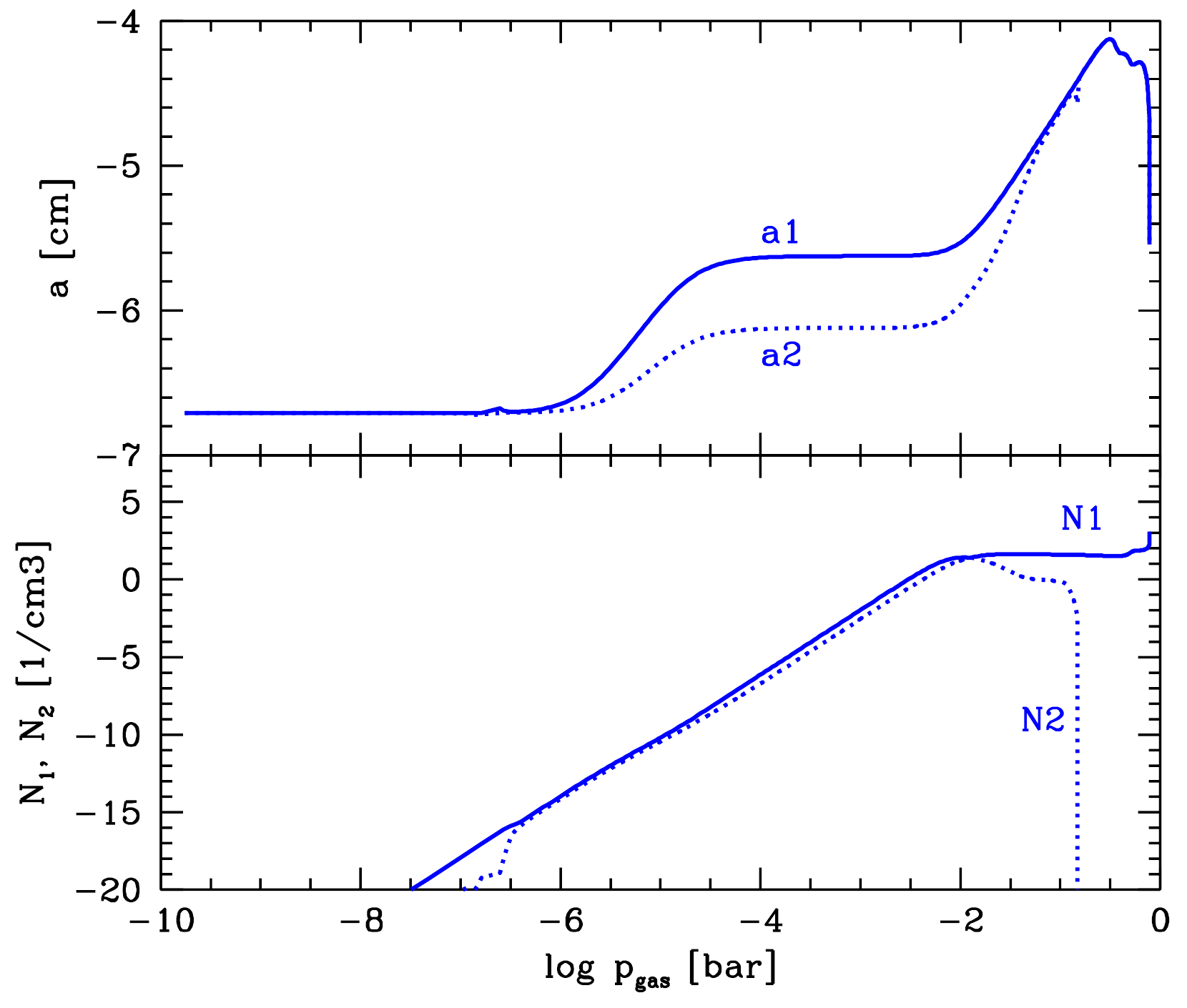}}
\resizebox{11cm}{!}{\includegraphics{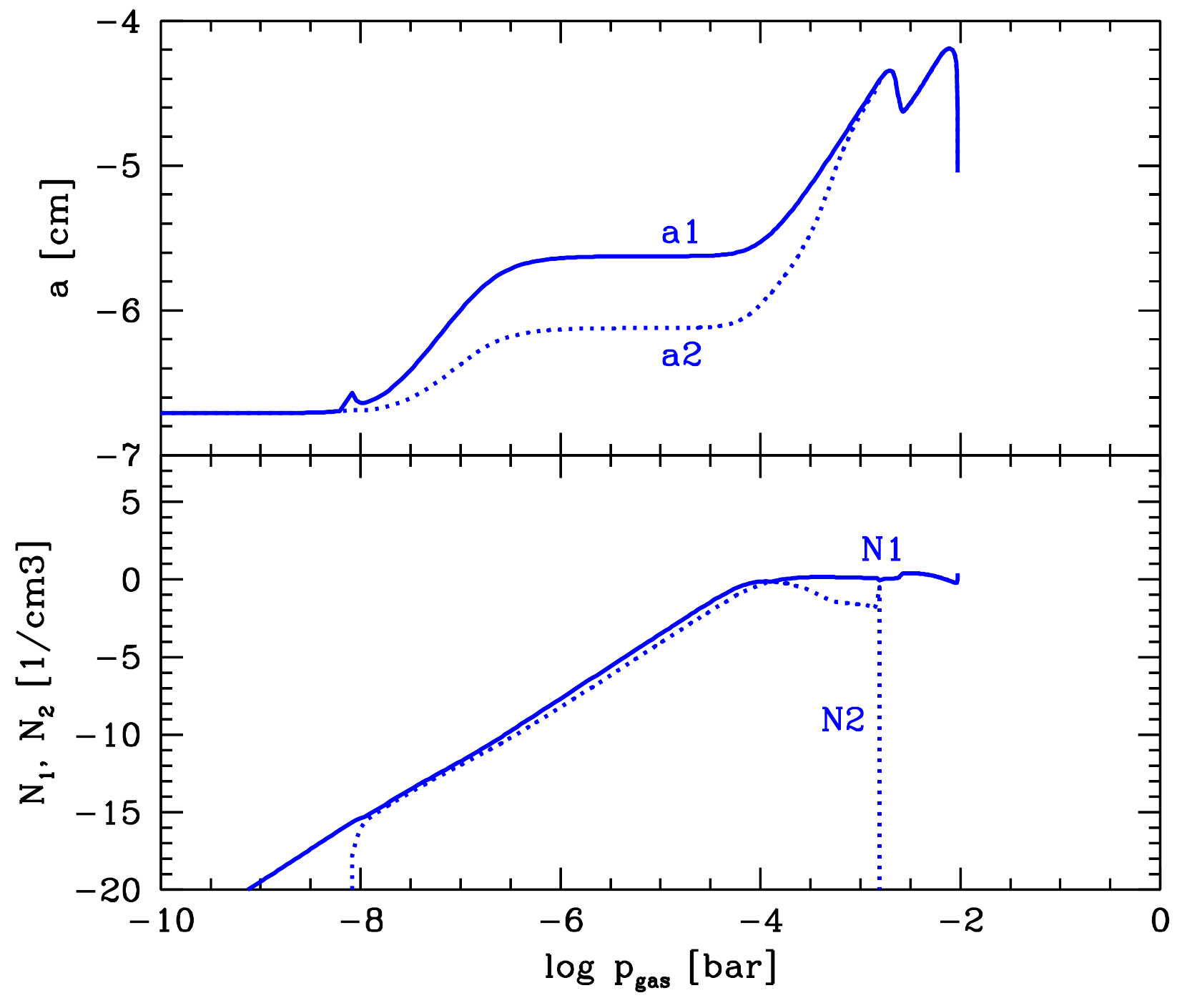}}
\caption{\small Height-dependent grain size distributions as result
from the {\sc Drift-Phoenix} atmosphere simulation. {\bf Top panels:} grain
sizes $a$,  {\bf Bottom panels:} number of grains $N$. $(a_1, N_1)$ and $(a_2, N_2$)
represent small and large particle ensemble in Eq.~\ref{eq:fvona}. Both model atmospheres are for T$_{\rm eff}=1600$K and solar metallicity. The {\bf top figure} represents a Brown Dwarfs with log(g)=5.0, and the {\bf bottom figure} a giant gas planet with log(g)=3.0.}
\label{fig:grain_size}
\end{center}
\end{figure}

The formation of clouds is the formation of solid particles or liquid
droplets in an atmosphere whose stratification is determined by
gravity, radiation, and convection.  Cloud formation in Brown Dwarfs
and giant gas planets starts with the formation of seed particles
(nucleation). This differs from the cloud formation process on
Earth (or solid planets in general) where dust grains that are carried
into the atmosphere from the planet's surface act as condensation
nuclei. In fully gaseous objects such as Brown Dwarfs or giant planets
those nuclei need to be built within the atmosphere itself.  Once these
seeds have formed, many materials are thermally stable and grow a
grain mantle by chemical surface reactions on the surface of each seed
particle. These compounds form patches of different, adjacent
materials on the grain surfaces as different solids can grow
simultaneously. Hence, these clouds are made of grains that are
composed of a mix of materials that depends on the local temperature
and pressure (Fig.~\ref{fig:Vd5},~\ref{fig:Vd3}, top panels). As a
consequence of these growth reactions, the mean particle size,
$\langle a \rangle $ (Fig.~\ref{fig:grain_size}, top panels),
increases.

Woitke \& Helling (2003) and Helling \& Woitke (2006) developed a
model describing such heterogeneous dust formation by homogeneous
nucleation, dirty growth, and evaporation. The formation of
TiO$_2$-seeds onto which solid silicate and oxide compounds grow is
considered.  Woitke \& Helling (2004) and Helling, Woitke \& Thi
(2008) applied this model to the formation of stationary clouds in
oxygen-rich\footnote{ {\it Oxygen-rich} refers to a gas that contains more oxygen then carbon, hence the gas phase is dominated by a diversity of oxygen-binding molecules compared to a carbon-rich gas. The models used here apply the solar element abundances which are oxygen-rich. } atmospheres including
gravitational settling, element depletion, and convective element
replenishment.  The atmosphere model {\sc Drift-Phoenix} (Dehn 2007,
Helling et al. 2008a,b; Witte, Helling \& Hauschildt 2009) couples
this detailed kinetic model of dust cloud formation with a radiative
transfer code (Hauschildt \& Baron 1999, Baron et al. 2003). We use
the output of the {\sc Drift-Phoenix} atmosphere simulations in this
paper.

 In {\sc Drift-Phoenix} the number of solids growing the mantle is
restricted to seven (TiO$_2$[s], Al$_2$O$_3$[s], Fe[s], SiO$_2$[s],
MgO[s], MgSiO$_3$[s], Mg$_2$SiO$_4$[s]) and to 32 surface reactions.
{\sc Drift-Phoenix} provides the local gas temperature T [K], the gas
pressure $p_{\rm gas}$ [dyn\, cm$^{-2}$], the maximum convective velocity $\rm
v_{\rm conv}^{\rm max}$ [cm\, s$^{-1}$], and dust quantities such as
the number density  of dust particles $n_{\rm d}$ [cm$^{-3}$] of mean grain
size $\langle a \rangle$ [cm] at each layer of the atmosphere. Because
of the definition of the dust moments, $\rho L_{\rm j} = \int_{\rm
V_l}^{\infty} V^{\rm j/3} f(V) dV$ ($V$ -- grain volume, $V_{\rm l}$ --
volume of smallest possible grain), the dust moments L$_{\rm j}$
($j=1, \ldots 4$) that result from a {\sc Drift-Phoenix} simulations
are used to derive a representative grain size distribution function
(for details see Appendix A in Helling, Woitke \& Thi 2008).  We consider in this
paper a double-peaked grain size distribution $f(a)$,
\begin{equation}
f(a) = N_1\delta(a-a_1) + N_2\delta(a-a_2) \label{eq:fvona}
\end{equation}
 to calculate the relative dust velocities
  (Sect~\ref{ss:dd_sed},~\ref{ss:dd_turb}) between grains of different
  sizes $a$ in each layer of the atmospheric cloud
  (Fig.~\ref{fig:grain_size}). Parameters $N_1$, $N_2$, $a_1$ and
  $a_2$ are determined such that the resulting dust moments reproduce
  the solution of the dust moment equations from the {\sc
  Drift-Phoenix} model atmosphere. Seed particles have the smallest
  size, $a_{\rm l}$, of all grains in the cloud and populate the
  uppermost cloud layers. As the cloud deck is defined as where the
  seed formation rate is maximum, small amount of very small particles
  form also above the cloud deck.

While the dust grains are forming and increasing in size they fall
under the influence of gravity into regions of increasing density
(gravitational settling, sedimentation).  The increasing gas density
results in faster grain growth, such that the grains increase their
size considerably over short distances, and in a decreasing
sedimentation velocity.
The relative velocities between dust and gas due to gravity can be
considered as the equilibrium drift velocity resulting from an
equilibrium between gravity and friction with the surrounding gas (see
Woitke \& Helling 2003).  The grains fall inward (i.e.  downward) and
they may reach a depth where the temperature is too high to sustain
their thermal stability.  For the example of a brown dwarf atmosphere,
at $T(p_{\rm gas}\approx 0.4$bar) in Fig.~\ref{fig:Vd5} (top panel)
the silicates evaporate, and instead, iron takes up $\approx 75\%$ of
the grain volume.  At even higher temperatures, i.e. higher gas
pressures in Fig~\ref{fig:Vd5} (top panel), the dust evaporates
completely which causes the decrease in grain size
(Fig~\ref{fig:grain_size}).  The same behaviour is found for the
example of a giant planet atmosphere but at lower pressure of $p_{\rm
gas}\approx 0.0025$bar . In both cases, the dust-to-gas ratios,
$\rho_{\rm d}/\rho_{\rm gas}$ (black solid line, bottom
Fig.~\ref{fig:Vd5}, ~\ref{fig:Vd3}) is highest just above the
evaporation zone where also most of the dust mass per cm$^2$ (red
dotted line) is located.

The dust influences the convective velocity $\rm v_{\rm conv}$
indirectly through the ``backwarming'' effect of its opacity on the
local temperature gradient of the atmosphere.  The backwarming changes
the convective behaviour of the atmospheres because of a flatter or
steeper local temperature gradient which then can suppress ($\delta
T/\delta r < \delta T/\delta r |_{\rm adiabatic}$) or initiate
($\delta T/\delta r > \delta T/\delta r |_{\rm adiabatic}$) convection
locally resulting in detached convection layers. The observation
of this effect is not new in model atmospheres and it can already be found in
the first papers in Brown Dwarf atmosphere modelling (e.g. Tsuji 2002;
Burrows, Sudarsky \& Hubeny 2006).  The convective energy transport
in the {\sc Drift-Phoenix} atmosphere simulations is considered in the
mixing-length approximation from which we derive a large-scale
convective velocity $\rm v_{\rm conv}$.  Helling (2005) showed that
dust formation also influences the local velocity field by radiative
cooling again due to the dust's large opacity. Hence, dust formation
can sustain turbulence which enhances the relative velocities of dust
grains on smaller scales than convection.  However, the influence of
the dust on the local turbulence spectrum is not taken into account in
the {\sc Drift-Phoenix} atmosphere simulation.

In this paper we use local quantities which have been
self-consistently evaluated in the {\sc Drift-Phoenix} atmosphere
simulation (except for the turbulence spectrum) describing the
formation of dust falling through a radiative and convective
atmosphere. We investigate to what extent collisional processes are
capable of ionising dust grains in clouds in atmospheres of Brown
Dwarfs and giant gas planets.  Figure~\ref{fig:grain_size} shows the
dispersion in grain sizes $a$ which produces a range of grain
velocities at each height in the atmosphere (compare also Helling,
Woitke \& Thi 2006, Helling \& Rietmeijer 2008).  Various processes
may influence the relative velocities between grains (Brownian motion,
convection, gravitational settling, turbulence). We are particularly
interested in whether collisional energies might then be large enough
to overcome the ionisation potential of an electron on the grain
surface, and hence if and how these processes can contribute to charge
cloud particles in the atmospheres of Brown Dwarfs and planets.  

We note that the local $(T, p_{\rm gas})$-structure can differ
amongst model families due to differences in cloud modelling
approaches (Helling et al. 2008).   Even model atmosphere simulations 
for hotter objects differ enough to impact  disk detection rates  (Sinclair et al. 2010). We are, however, confident that the {\sc Drift-Phoenix} model atmospheres are a very reasonable attempt as
the comparison to observations in the {\it DwarfArchieve} demonstrates (Witte, Helling \&
Hauschildt 2011).

 We perform our study for a brown dwarf and a gas planet
case. We choose two solar metallicity {\sc Drift-Phoenix} models for a
rather cool atmosphere of T$_{\rm eff}=1600$K. The high-gravity case
($\log$g=5.0) serves as an example for a brown dwarf atmosphere and the
low-gravity case ($\log$g=3.0) is the example for a gas planet
atmosphere. This brown dwarf atmosphere model was also used in Helling
et al.~(2010).

\section{Work functions for mixed-material dust}\label{ss:wf}

 The release of an electron from the grain's surface requires that it
overcomes its  ionisation potential, which is the
energy needed to extract an electron from a surface and transport it
to infinity.
We assume  this to be the work function in metals or the contact
potential for insulators (see discussion in Poppe \& Schr\"apler
2005). The values of the work function are mostly given for pure
atomic materials in the literature (Desch \& Cuzzi 2000, Kopnin et
al. 2004) and range from $\approx 2 \ldots 6$eV
(Fig.~\ref{fig:col_energy}). However, our grains are made of
well-mixed solid materials forming from an oxygen-rich gaseous
environment, and they are only partially dielectric.  Rosenberger
(2001) states that the work function of multi-material grains is
typically lower than the ionisation potential of pure atomic
materials.  We therefore conclude that the work function for grains of mixed materials is not well constrained, and we indicate the potential range of work functions in Fig.~\ref{fig:col_energy} (orange bar). The upper value may be somewhat overestimated leaving space for processes that we did take into account.
Ideally, these electrons
would be free to move around, but grains are observed to
preferentially produce surface charges on their collisional partners
rather than free electrons. We start our investigation with
considering multi-electron emission due to dust collisions  which may remain on the grain surface or escape if their energy is large enough. We see
this process  somewhat  in analogy to secondary electron emission where more then
one electron could be released during the collision with an electron
or ion. During such collisions,  backward scattering is more efficient
in lattices of semi-conductor materials or insulators compared to
metals  which is of interest because our cloud model predicts
particles made of a mix of materials. The secondary electron emission
coefficient ranges from 2.4 for MgO[s] to 4.6 for Al$_2$O$_3$[s]
(NaCl[s]: 6) and is highest for mixed materials (Ag-Cs$_2$O-Cs: 8;
Niedrig 1992, p359). We therefore adopt these secondary electron
emission coefficients, i.e. the number of free electrons produced per
collision, as guidance for mixed materials.  We further note in
analogy to the secondary electron emission that an increased
collisional energy will not necessarily cause a continuously
increasing number of electron releases because the impact may affect
deeper electrons in the solid which require a larger energy.

\section[]{Collisional ionisation}\label{s:colion}

We study ionisation events by collisions inside the dust cloud layers
that form in Brown Dwarfs and planetary atmospheres. We investigate
limiting cases in order to study if grains are charged in substellar clouds and
hence, if they can act as seeds for other, more powerful mechanisms
like e.g. electron avalanche processes. We compare the collisional energies
to the ionisation energies of the dust grain surface. Three collision
mechanisms are studied:\\*[-0.5cm]
\begin{tabbing}
--  dust-gas collisions due to gravitational settling\\
-- dust-dust collisions by differential sedimentation of cloud particles\\
-- dust-dust collisions due to turbulence
\end{tabbing}

 We discuss the effect of collisional processes only. During an
elastic collision, the kinetic energy remains constant as the internal
energy of the colliding bodies does not change. But the internal
energy changes in the case of an inelastic collision e.g. due to
chemical processes, and hence, the kinetic energy of the colliding
grains must change. Such inelastic collision could cause
fragmentation, erosion, sticking, deformation or evaporation
(G\"uttler et al. 2010), besides the electron release discussed
here. The occurrence of these processes, which are not considered in
this paper, depends on the grain mass and relative velocity.


\subsection{Collisional energy and work function}
The collisional energy is calculated for two-body collisions
which can either be dust-gas or dust-dust collisions. This
collisional energy is then the kinetic energy for these two-particle
collisions,
\begin{equation}
\label{eq:Ecol}
E_{\rm col}=\frac{1}{2}m_{\rm red}{\rm v}_{\rm rel}^2.
\end{equation}
Here, ${\rm v}_{\rm rel}$ is the relative velocity between the
collisional partners and $m_{\rm red}$ is the reduced mass for a two
particle collision. Thus for dust-gas collisions ({\rm dg}):
\begin{equation}
m_{\rm red, dg}=\frac{m_{\rm d} m_{\rm g}}{m_{\rm d}+m_{\rm g}},
\end{equation}
and for dust-dust encounters ({\rm dd'}):
\begin{equation}
m_{\rm red, dd'}=\frac{m_{\rm d} m_{\rm d'}}{m_{\rm d}+m_{\rm d'}}
\end{equation}
where $m_{\rm d}=\frac{4}{3}\pi a^3 \rho_{\rm d}$ is the mass of a
dust grain,  $m_{\rm d'}$ the mass of the colliding grain,  and $m_{\rm g}=\bar{\mu}$  the mean molecular weight in [amu] of the gas .

Clearly, the collisional energy depends on the relative
velocity of the collision. We therefore consider the above mentioned
three possible types of collision separately in the following
sections. 

\subsubsection{Dust-gas collisions due to gravitational settling}\label{ss:dg}

In Sec. 2 we have seen that, once the dust starts forming, each grain
falls through its gaseous environment under the influence of
gravity and friction. The equilibrium sedimentation velocity between gravitation
and frictional force during sedimentation is reached quickly, and
therefore, we can apply the equilibrium drift concept (Woitke \&
Helling 2003) to calculate the sedimentation velocity, ${\rm v}^{\rm
sed}(a)$.  The dust sedimentation velocity is relative to the gas
phase and depends on the grain size $a$. We restrict this
consideration to the large Knudsen number case (for a
grain moving through gas) to be consistent with the
quasi-static dust model applied in Witte, Helling \& Hauschildt (2009)
whose {\sc Drift-Phoenix} model atmosphere results we will use.  We,
however, distinguish between the subsonic and the supersonic case
where $c_T$ is the local sound speed:

\begin{eqnarray}
\label{eq:vsed1}
{\rm v}^{\rm sed}(a)=\frac{\sqrt{\pi}g \rho_{\rm d} a}{2 \rho_{\rm gas} c_T} &&\qquad  {\rm v}^{\rm sed} < c_T\\
\label{eq:vsed2} {\rm v}^{\rm sed}(a)=\sqrt{\frac{4 a \rho_{\rm d} g}{3
\rho_{\rm gas}}} \cdot c_T &&\qquad  {\rm v}^{\rm sed} > c_T\,.
\end{eqnarray}
The dust grains move with velocities higher than the sound speed 
in the very outer layers of the atmosphere due to the very low densities of the gas phase in these regions. 
 A small-scale turbulent fluid field would influence the sedimentation velocity
such that grains have more time and grow to larger sizes.

\subsubsection{Dust-dust collisions due to differential sedimentation}\label{ss:dd_sed}

Differential sedimentation due to different grain sizes provides
a source of collisions amongst the dust particles themselves.
Applying the above representation of the grain size
distribution function (Eq.~\ref{eq:fvona}) in each layer of the
atmospheric cloud  to Eq.~\ref{eq:vsed1} and Eq.~\ref{eq:vsed2}
allows us to calculate relative sedimentation velocities, ${\rm
v}^{\rm sed}_{\rm rel}(a)$, between grains of different sizes in
each atmospheric layer caused by differential sedimentation.
 In order to illustrate the range of collisional velocities we calculate
the relative velocities between the lower and upper grain size
limits $a_1$ and $a_2$:

\begin{eqnarray}
\label{eq: vrela1a2}
{\rm v}^{\rm sed}_{\rm rel\,
a_2,\,a_1}(a)&=&{\rm v}^{\rm sed}(a_1)-{\rm v}^{\rm sed}(a_2)\,.
\end{eqnarray}
At high altitudes where p$_{\rm
gas}<10^{-6}{\rm bar}$, only nucleation is possible. Therefore, the
variation in grain sizes is zero and all grains have the same
sedimentation velocity at a given height, hence ${\rm v}^{\rm
sed}_{\rm rel}(a)=0$.  This behaviour is quite general
 as Witte, Helling \& Hauschildt (2009) show
in their Fig.~2 for a whole grid of substellar atmosphere models.
The
greatest relative velocity is achieved in the region of efficient
growth where the grain size differences can be largest.  However, the
relative velocity drops with higher pressures because the $\rho_{\rm
gas}$-dependency dominates if $a\approx$const because ${\rm v}^{\rm
sed}(a)\sim a/\rho_{\rm gas}$ (or ${\rm v}^{\rm sed}(a)\sim
\sqrt{a/\rho_{\rm gas}}$ for ${\rm v}^{\rm sed} > c_T$; Eqs.~\ref{eq:vsed1},~\ref{eq:vsed2}).

\subsubsection{Enhanced dust-dust collisions in a turbulent gas}\label{ss:dd_turb}

A further velocity component leading to inter-dust collisions results
from a turbulent fluid field. The role of turbulence in the
atmospheres of substellar objects such as Brown Dwarfs and planets has
been studied in Helling et al. (2001). Convective mixing and gravity
waves are mechanisms to drive a turbulent fluid field in a substellar
atmosphere that otherwise would be damped by the viscosity of the
atmospheric gas. Whereas the frictional coupling between the dust and
gas is important in determining the terminal velocity during
sedimentation, friction can also cause these particles to couple to
the turbulent component of the gas velocity. Whether the dust grains
couple to the turbulent motion of the gas depends on the relation
between the frictional time scale, $\tau_f$, and the turnover time,
$\tau_t$, of a turbulent eddy (e.g. V{\"o}lk et al. 1980).  Particles
of a given size accelerate to their terminal velocity over a
frictional timescale $\tau_f$ (i.e., the grain's acceleration time
scale $\tau_{\rm acc}$ in Sect. 2.5 in Woitke \& Helling 2003). They
can, therefore, couple to the velocity field of those turbulent eddies
whose turnover time $\tau_t$ is larger than $\tau_f$ ($\tau_t >
\tau_f$). The frictional coupling time is for the subsonic and the
supersonic case (Woitke \& Helling 2003),
\begin{equation}
\label{eq:tauf} \tau_f =  \left\{\begin{array}{ll}
  \frac{2\sqrt{\pi}\rho_{\rm d}a}{3\rho_{\rm gas} c_T}
  & \quad \,{\rm v}^{\rm sed} \ll c_T \\[1.3ex]
  \frac{2a\rho_{\rm d}}{3\rho_{\rm gas} {\rm v}^{sed}}
  & \quad \,{\rm v}^{\rm sed} \gg c_T. \end{array}\right.
\end{equation}
For homogeneous, isotropic turbulence the energy
dissipation rate $\epsilon$ is constant for all scales (Kolmogoroff's
theory), and one finds from dimensional arguments that inside the
inertial range
\begin{equation}
\epsilon = C_1 \frac{u^3}{l},
\label{equ:dissip}
\end{equation}
 where $u$ is the fluid velocity associated with a scale $l$. The
 constant $C_1=0.7$ (Jimenez et al., 1993).  We determine the locally
 constant energy dissipation rate $\epsilon$, for the largest scale of
 maximum kinetic energy assuming $l_{\rm max}=H_p(r)/10$ (pressure
 scale height $H_p(r)\approx 10^4 \ldots 10^6$cm depending on radial distance $r$) and adopt the
 maximum vertical RMS velocity, $u$
 as proposed in Freytag et al. (2010, using their Eq. 5 with $\log
 V=\log u_{\rm max}$),
\begin{equation}
\log u_{\rm max}(r) = \log {\rm v}_{\rm conv}^{\rm max} + \log r_v - (\log p_{\rm gas} - \log p_{\rm gas}^{\rm max})/(H_v/H_p)
\label{equ:u}
\end{equation}
 with ${\rm v}_{\rm conv}^{\rm max}$ the maximum convective
velocity and $p_{\rm gas}^{\rm max}$ the gas pressure at this maximum
velocity taken from our {\sc Drift-Phoenix} atmosphere simulation.
Equation~(\ref{equ:u}) parametrises the velocity of the large scale
that drives the turbulent fluid field in the atmosphere according to
the local thermodynamic conditions. Such large scale motions can be
either convection or gravity waves according to Freytag et
al. (2010). Freytag et al. (2010) derive parametrisations for
$H_v/H_p$ (velocity scale height of the wave amplitude; their Eq. 3)
and $r_v$ (the ratio of maximum convection energy to wave amplitude =
'mixing efficiency', their Eq. 4) from their grid of 2D convection
models for substellar atmospheres. In this parametrisation, $H_v/H_p$
and $\log r_v$ are constant for a given set of stellar parameters. Hence, we
find $H_v/H_p=2.855$ and $\log r_v = -1.348$ for our cases of T$_{\rm
eff}=1600$K (solar metallicity) from their Eqs.~3, 4.  With ${\rm
v}_{\rm conv}^{\rm max}=4.5\cdot 10^3$ cm/s and $p_{\rm gas}^{\rm
max}=2.0\cdot 10^7$ dyn/cm$^2$ (=20\,bar) for $\log(g)=5,0$, and
${\rm v}_{\rm conv}^{\rm max}=1.7\cdot 10^4$ cm/s and $p_{\rm
gas}^{\rm max}=3.0\cdot 10^5$ dyn/cm$^2$ for  $\log(g)=3,0$,
Eq.~(\ref{equ:u}) reduces to
\begin{eqnarray}
\label{equ:u1}
\log u_{\rm max}(r) = 4.862 - \frac{\log p_{\rm gas}}{2.855} && (\mbox{brown dwarf},\log(g)=5.0 ) \\
\log u_{\rm max}(r) = 4.800 - \frac{\log p_{\rm gas}}{2.855} && (\mbox{giant planet},\log(g)=3.0)
\label{equ:u2}
\end{eqnarray}
Hence, the maximum vertical RMS velocity has no strong dependence on the surface gravity.
The energy dissipation rate $\epsilon$ (Eq.~\ref{equ:dissip}) depends on the atmospheric
pressure $p_{\rm gas}$, and it therefore changes globally but it is constant
in each atmospheric layer $\Delta h_{\rm i}$ (see Sect.~\ref{ss:pecoll}).

The eddy turn-over time, i.e.  the
 correlation time of turbulent fluctuations, is
\begin{equation}
\tau_{\rm t} = \frac{l}{u}\sim\left(\frac{l^2}{\epsilon}\right)^{1/3},
\label{equ:turnover}
\end{equation}
and it depends on the locally characteristic scale $l$ (eddy size).
Since $\tau_{\rm t}$ determines the lifetime of an eddy, smaller
eddies have a shorter lifetime than larger eddies for a constant
energy dissipation rate $\epsilon$. Hence, if we know $\epsilon$ and
$l$, we can determine $\tau_{\rm t}$.

\begin{figure}
\begin{center}
\resizebox{11cm}{!}{\includegraphics{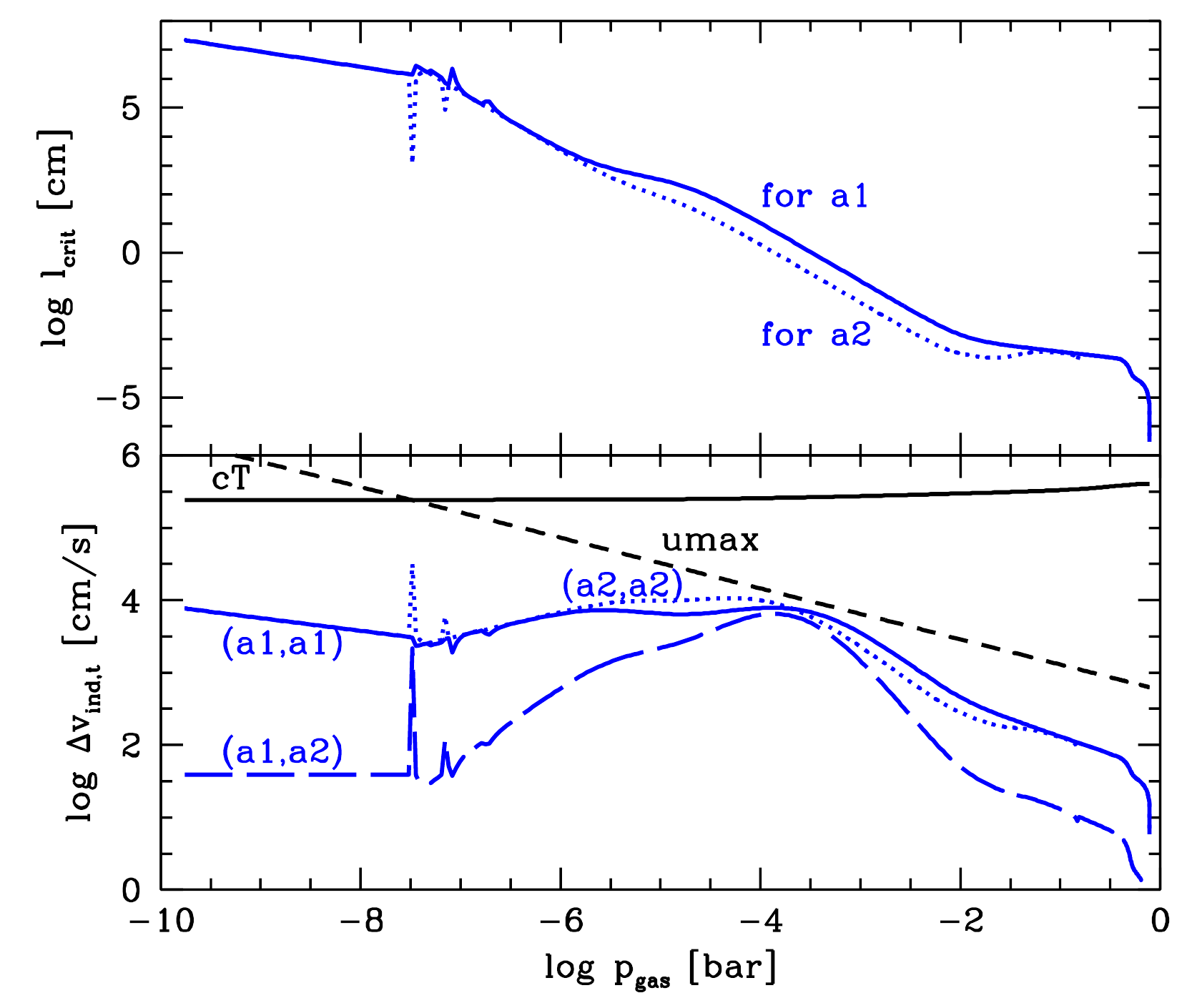}}
\resizebox{11cm}{!}{\includegraphics{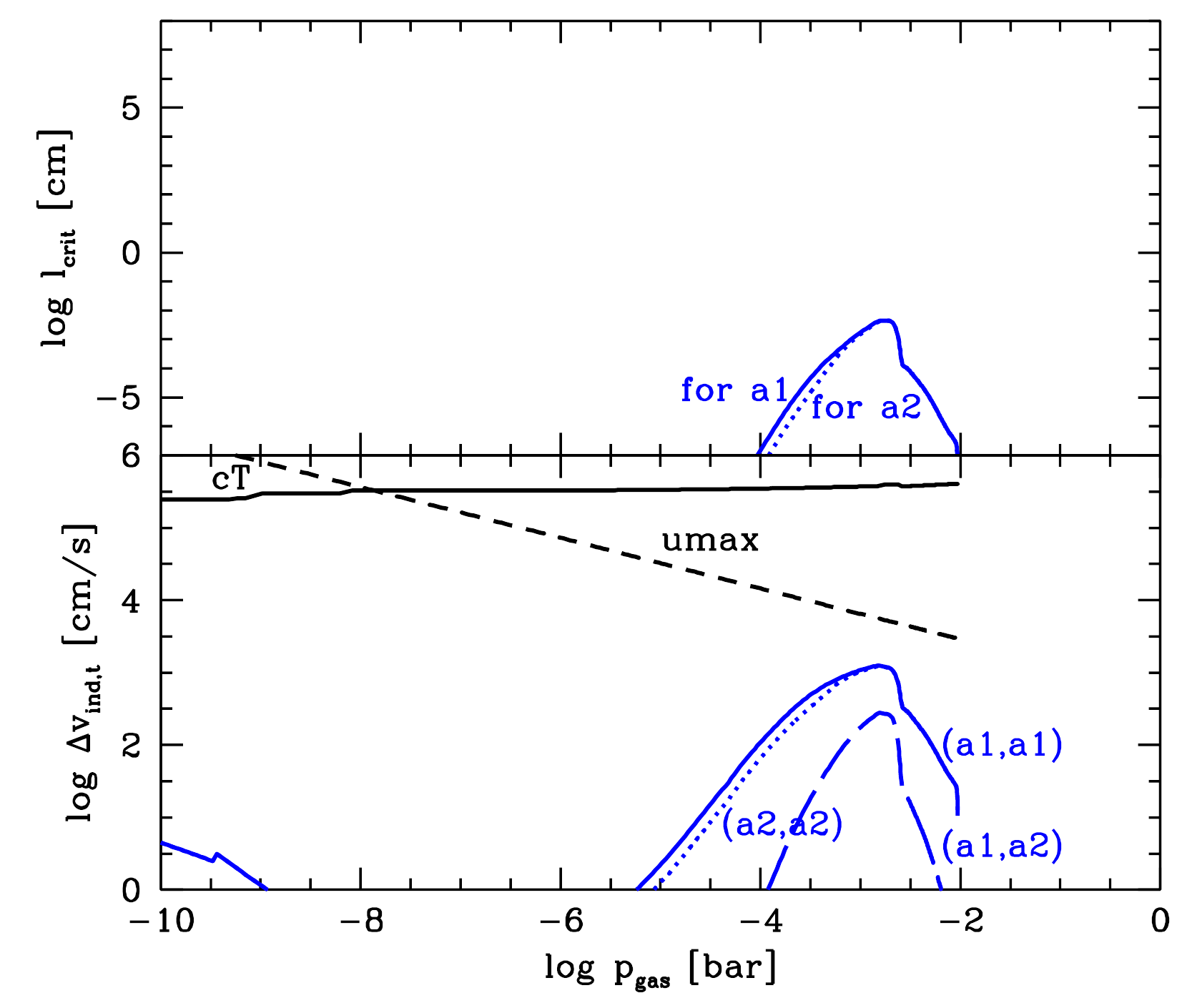}}
\caption{\small {\bf Top panel:} Critical length scale for which
$\tau_{\rm f}=\tau_{\rm t}$. {\bf Bottom panel:} Turbulence induced
relative inter-grain velocity component ${\rm v}_{\rm ind,t}$ applying
$l=1$cm (Eq.~\ref{equ:turnover}). The local sound speed c$_{\rm T}$
and the large-scale gas velocity, $u_{\rm max}$ which drives the
turbulence in the gas are over-plotted. Both model atmospheres are for
T$_{\rm eff}=1600$K and solar metallicity. The {\bf top figure}
represents a Brown Dwarfs with log(g)=5.0, and the {\bf bottom figure}
a giant gas planet with log(g)=3.0.}
\label{fig:v_turb_ind}
\end{center}
\end{figure}

Whether and how the turbulent gas motion influences the dust dynamics
depends on the relation between $\tau_f$ and $\tau_t$. Under certain
conditions the turbulent gas interacts with the dust grains such that
the dust grains acquire a relative inter-grain velocity component
among them.  In order to quantify the influence of turbulence on the
dust dynamics, we use the formalism
deduced by Morfill (1985)  to calculate the turbulence-induced relative
(drift) velocity between the grains in the cloud, $\Delta {\rm v}_{\rm ind,t}$:
\begin{eqnarray}
\Delta {\rm v}_{\rm ind,t} \simeq \nonumber \\
\langle \delta
{\rm v}_g^2\rangle^\frac{1}{2}\big[(1+\frac{\tau_{f1}}{\tau_t})^{-1}+(1+\frac{\tau_{f2}}{\tau_t})^{-1}\big)\nonumber \\
\label{eq:vindt}
-2\big(\frac{1}{(1+\frac{\tau_{f1}}{\tau_t})(1+\frac{\tau_{f2}}{\tau_t})}\big)\big].
\end{eqnarray}
Equation~\ref{eq:vindt} simplifies for grains of similar sizes ($\tau_{f1}\sim\tau_{f2}$) to
\begin{equation}
\label{eq:vindt2}
\Delta {\rm v}_{\rm ind,t} \simeq \langle \delta {\rm v}_g^2\rangle^\frac{1}{2}
\frac{\sqrt{\frac{2\tau_f}{\tau_t}}}{1+\frac{\tau_f}{\tau_t}}.
\end{equation}Ä
Following Morfill (1985) we represent $\rm \langle \delta {\rm
v}_g^2\rangle^\frac{1}{2}$, which is any systematic velocity
component, by Eq.~\ref{equ:u} or Eqs.\ref{equ:u1}/\ref{equ:u2}.  To
calculate $\tau_t$, we consider a typical eddy scale of $l=1$cm to
represent small-scale end of the turbulence spectrum.  This
choice is guided by the critical length scale $l_{\rm cri}$ for which
$\tau_f=\tau_t$ (Fig.~\ref{fig:v_turb_ind}, top panels) i.e. when the
momentum exchange between gas and dust is most efficient, and
consequently the turbulence-induced inter-grain velocity component is
highest. We can argue with Eq.~\ref{equ:turnover} that the eddies have
already decayed before the grain fictionally couple to the gas
($\tau_f>\tau_t$) for eddy sizes $l<l_{\rm cri}$. The grains
equilibrate with the gas, hence they move together with the turbulent
eddies only if $\tau_f<\tau_t$ for which $l>l_{\rm cri}$ follows. 
Figure~\ref{fig:v_turb_ind}  demonstrates that the behaviour is rather diverse in the Brown Dwarf case for our choice of  $l=1$cm, pointing to a multi-scale problem that only very approximately can be described by one pre-chosen scale (see also Helling 2005). In the planetary atmosphere case,  our choice of  $l=1$cm suggests that all grains are frictionally coupled. If $l$ is treated in more detail, the impact of turbulent motions on dust-dust collisions will be larger  in the planetary atmosphere because $l_{\rm cri}<1$cm.  Note, however, that grains will decouple from the gas due to gravity
on the largest scales in the atmosphere.

Figure~\ref{fig:v_turb_ind} (bottom panels) shows the
turbulence-induced relative (drift) velocity ${\Delta \rm v}_{\rm
ind,t}$ for collisions of grains of different sizes (dashed line,
Eq.~\ref{eq:vindt}) and for grains of the same size (solid and dotted
line, Eq.~\ref{eq:vindt2}) over the whole atmospheric pressure
range. The turbulence driving velocity, $u_{\rm max}(p_{\rm gas})$, is
plotted in comparison. There is a maximum of ${\Delta \rm v}_{\rm
ind,t}$ in both atmosphere cases, brown dwarf and gas planet, whereas
${\Delta \rm v}_{\rm ind,t}$ clearly decreases towards higher
pressures. In these inner atmospheric regions, $\tau_f$ decreases
because the gas density increases, and therefore the dust grains can
adjust more quickly to the gas motion.

\subsection{Energy release during dust-gas and  dust-dust collisions}\label{sss:en}

\begin{figure}
\begin{center}
\resizebox{11cm}{!}{\includegraphics{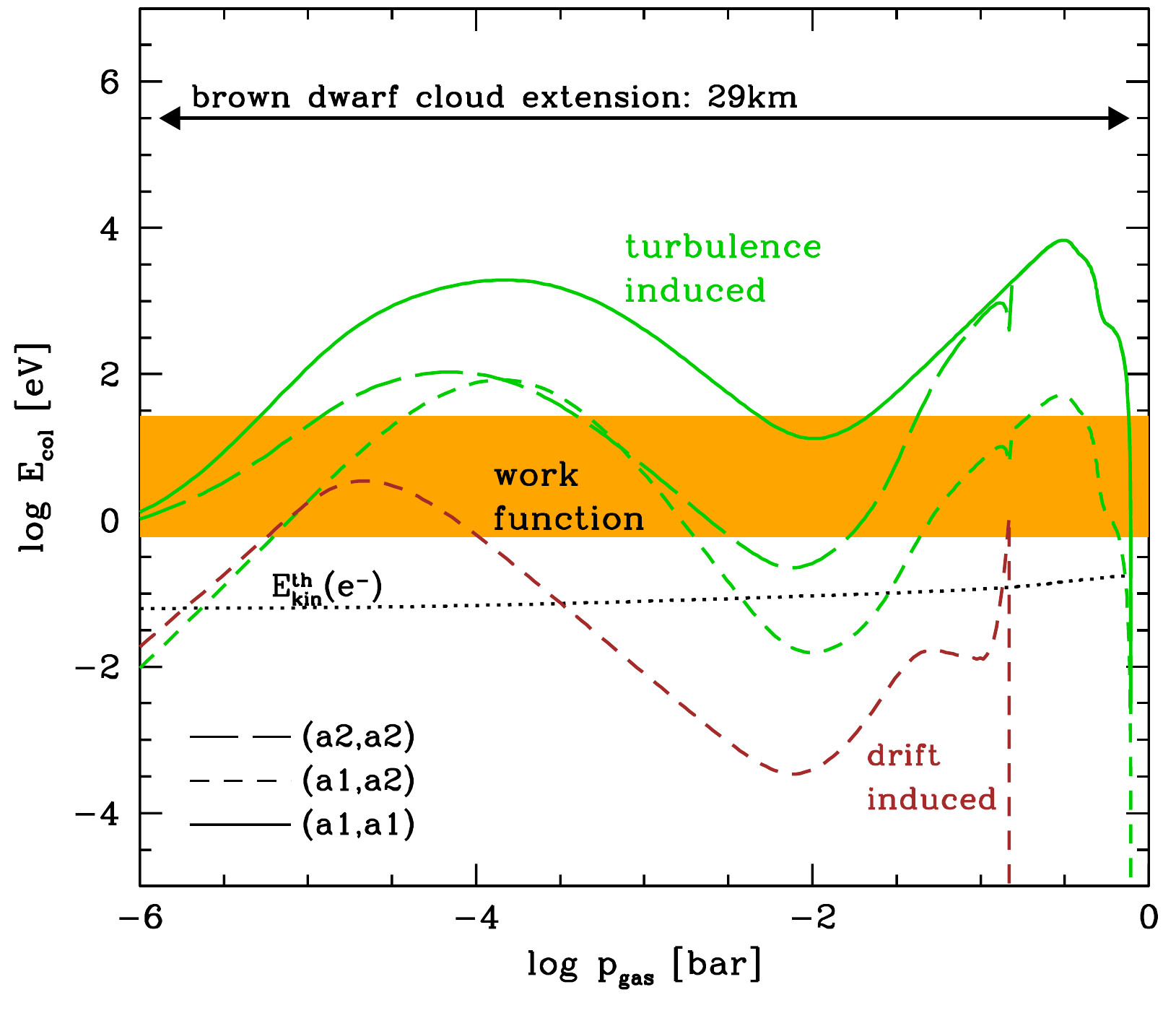}}
\resizebox{11cm}{!}{\includegraphics{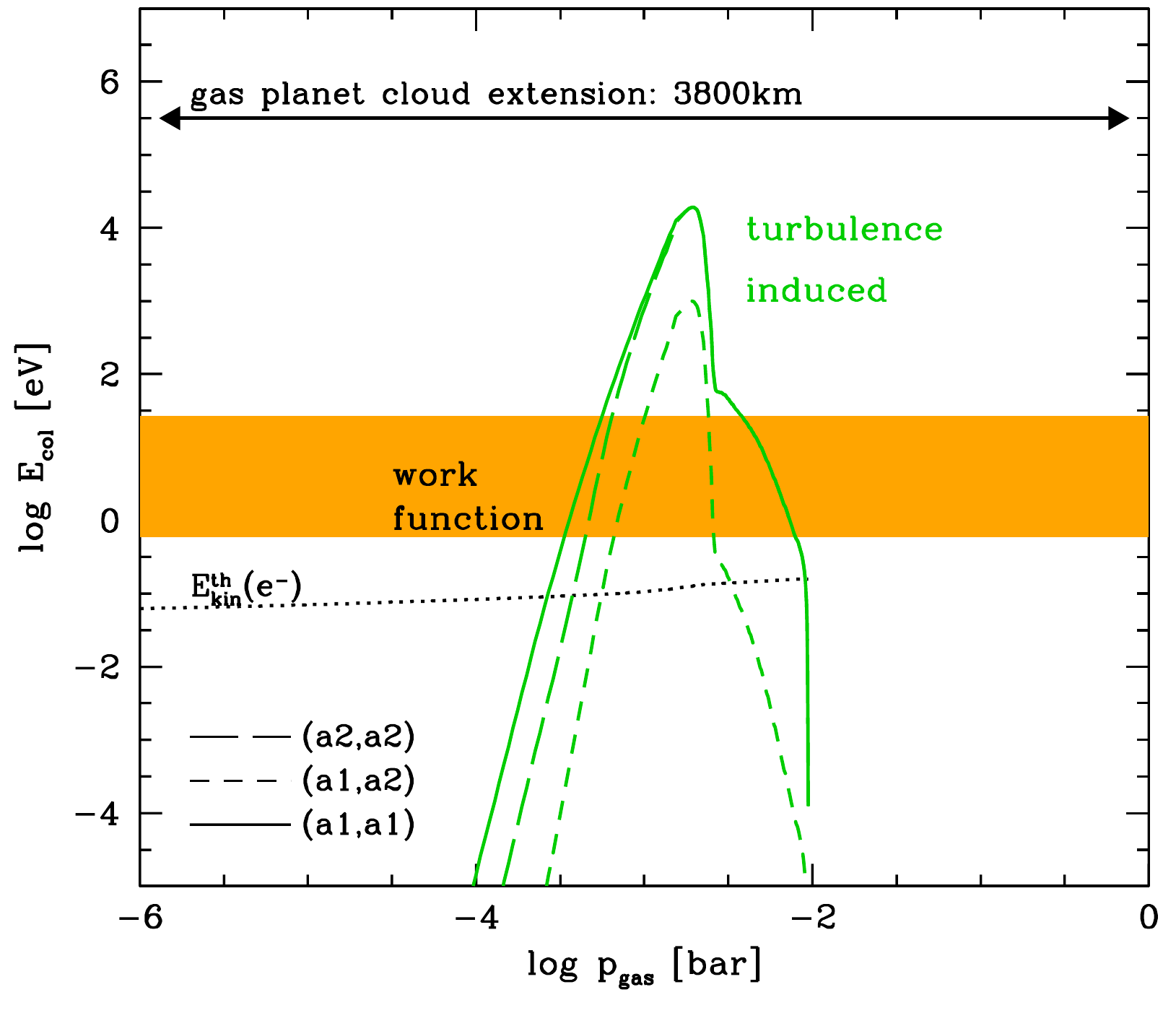}}
\caption{\small Collisional energies [eV] due to turbulent dust-dust
encounters (green) and due to dust-dust encounters (brown) during
gravitational settling. The electrons thermal kinetic energy, $E^{\rm
th}_{\rm kin}(e^-)$, is overplotted as black dotted line. The range of plausible dust work
function is shown as orange bar. Both model atmospheres are for
T$_{\rm eff}=1600$K and solar metallicity. The {\bf top figure}
represents a Brown Dwarfs with log(g)=5.0, and the {\bf bottom figure}
a giant gas planet with log(g)=3.0.}
\label{fig:col_energy}
\end{center}
\end{figure}

The collisional energies according to
Eqs.~\ref{eq:Ecol},~\ref{eq:vsed1},~\ref{eq:vsed2}, and Eqs.~\ref{eq:
vrela1a2}, and Eq.~\ref{eq:vindt2} are shown in
Fig.~\ref{fig:col_energy}, together with the ionisation potential
(work function) for a mix of  solid materials as it is expected
for a grain surface based on our dust formation model.  
We explore the results of our considerations for two solar metallicity
{\sc Drift-Phoenix} model atmospheres : T$_{\rm eff}=1600$K for
$\log(g)=5.0$ (brown dwarf) and $\log(g)=3.0$ (planet).

Figure~\ref{fig:col_energy} shows that the collisional energies can
change by  1-5 orders of magnitude  throughout the cloud layer of
the brown dwarf and even more within the planetary atmosphere. This
change is partly directly due to the variation in grain mass which in turn
depends on the local thermodynamic conditions in the atmosphere, and
partly due to changing local conditions, such as the relative grain
velocity, temperature, and pressure.

The collisional energies due to dust-gas collisions (not shown) are
{\it not} large enough to provide the ionisation energy, and hence,
this kind of collisions are negligible for grain ionisation.  Dust-dust
collisions between grains of different sizes due to differential
sedimentation (``rain out'', short dashed brown line) reach the
ionisation potential for grains only at rather low gas pressures
($p_{\rm gas}=10^{-5}\,\ldots\,10^{-4}$bar) in the brown dwarf atmosphere. Comparing
Fig.~\ref{fig:col_energy} with Fig.~\ref{fig:grain_size} shows
that this interval is related to strong changes in grain sizes due to
grain growth and grain evaporation.

For turbulence-induced collisions between grains of similar sizes
(green solid line, Fig.~\ref{fig:col_energy} ) produce the largest
collisional energies. Most of the turbulence induced inter-grain
collisions have collisional energies larger than the ionisation
potential and are therefore most efficient in ionising grains inside
the clouds in substellar atmospheres.

 Our example of a giant gas planet atmosphere demonstrates that
dust-dust collisions are only sufficient to overcome the dust
material's ionisation energies in the inner and densest part of the
cloud, and only of turbulence amplifies the collisional process. This
effect will amplify if the inter-grain velocities are larger due to
a more efficient frictional coupling then assumed in this
paper. However, the geometrical height of the cloud covered is larger than in
a high-gravity brown dwarf atmosphere (Fig.~\ref{fig:col_energy}).

\subsection{Electron pressure equivalent due to dust-dust collisions}\label{ss:pecoll}

 In the first paper of this series, we argued that charged grains
can act like small capacitors and initiate electron avalanche
processes that produce large numbers of free electrons. If many of
these events superimpose and enough electrons leak out of the streamers, the degree of gas ionisation could be
increased to such an extent that the cloud might couple to a large scale
magnetic field. In this paper, we have demonstrated that the cloud
particles  can be charged in substellar and planetary atmospheres on the
basis of turbulence induced  inter-grain collisions alone. The next step is to
quantify the number of electrons that
potentially can be produced by these collisions in the  whole cloud. This would also
allows to investigate if inter-grain collisions have the potential to
produce enough charges to allow a coupling with a large
scale magnetic field in the case that these charges are released into
the gas phase where they have a greater mobility.

The energies produced by dust-dust collisions are larger than the
ionisation energy by several orders of magnitude in most of the brown
dwarf cloud volume or a fraction of it in gas planets if the cloud is
turbulent.  Such high energies which also exceed the thermal electron
energy (black dotted line, Fig.~\ref{fig:col_energy}), suggest that
electrons could escape from the grain surfaces into the gas. Whether
the electrons affected by dust-dust collisions remain within the dust
phase or whether they indeed escape from the grain surface depends on
their kinetic energy gained during the collisions and on the
electronegativity of the surrounding gas phase constituents.  In
the case of destructive grain processes, the potentially
charge-carrying dust surface increases and charges
could escape more easily into the gas as evaporation processes are
faster.  Many more mechanisms can influence this process (see introduction to Sect.~\ref{s:colion}) in addition to profound uncertainties in the microphysics of dust charging.

Desch \& Cuzzi (2000) consider frictional charge transfer between
grains of different size. These authors assume grains of homogeneous
composition, i.e. insulator grains vs. metal grains, which is unlikely
in oxygen-rich astrophysical gases, at least in the case of substellar
atmospheres. Figures~\ref{fig:Vd5},~\ref{fig:Vd3} show that grains are composed of a
mixture of metal and insulator materials over a large fraction of the
cloud's extension.  Hence, a charge-conservation model for homogeneous
particles can not straight forwardly be applied to substellar
atmospheres. Further, the collisional energies due to turbulence induced inter-grain velocities exceed the level of ionization in some regions by a
few orders of magnitude. That implies that the number of electrons
which are supposed to be exchanged between dust grains by
triboelectric effects according to Desch \& Cuzzi might eventually be
released off the dust phase, having gained a certain amount of
kinetic energy during the collision. We therefore approximate these
possibly free electrons in the form of a non-thermal electron pressure
to quantify the effect.

We consider the frequency of collisions between dust particles of
radius $a$ and dust grains of another size $a^{\prime}$
(\footnote{This frequency will increase or decrease depending
on the charge number of the grain itself in a distance smaller then
the Debye length.  Diver \& Clark (1996) and Stark et al. (2006)
showed that the grain's electric field will be stronger if the grain 
departs from spherical symmetry.}). We denote the size distribution of
these collisional partners as $f(a^{\prime})$ and write the collision
frequency $\nu_{\rm col}$ as,
\begin{equation}
\label{eq:nucolint}
\nu_{\rm col}=\pi \int_{\rm a_l}^{\infty} a_{\rm red}^2\,{\rm v}_{\rm rel}(a,a^{\prime})\,f(a^{\prime})\, da^{\prime}
\end{equation}
 with $a_{\rm red}= a\,a^{\prime}/(a+a^{\prime})$ the reduced grain
radius and $a_{\rm l}$ is the radius of the smallest grain possible (for details see
Sect.~5 in Woitke \& Helling 2003).  Assuming a delta-function-like
representation of the grain size distribution function $f(a^{\prime})=
n^{\prime}_{\rm d}(a^{\prime}) \delta(a-a^{\prime})$,
Eq.~\ref{eq:nucolint} simplifies to
\begin{equation}
\nu_{\rm col}=\pi\,  a_{\rm red} ^2\,{\rm v}_{\rm rel}(a,a^{\prime})\,n_{\rm d}^{\prime}(a^{\prime}) .
\end{equation}
Here, $n_{\rm d}$ [cm$^{-3}$] is the number density of the
collisional partner of the grain of size $a$.  The relative velocity
${\rm v}_{\rm rel}(a,a^{\prime})$ between grains of different sizes
given by Eq.~\ref{eq: vrela1a2},  Eqs.~\ref{eq:vindt} and~\ref{eq:vindt2}. From Fig.~\ref{fig:col_energy} we conclude that only Eqs.~\ref{eq:vindt} and~\ref{eq:vindt2} are relevant so far.

We consider dust grains of size $a$ passing through  consecutive  cloud fractions
of thickness $\Delta h_{\rm i}(T,p)$   ($\Sigma_{\rm i}\,\Delta h_{\rm i}$ =
geometrical cloud extension $\Delta H_{\rm cloud}$, Fig.~1 in Helling
et al. 2010) colliding with other grains at a certain height in the atmosphere.
Note that $\Delta h_{\rm i}$ will increase with decreasing log(g) as
the geometrical extension of the cloud, $\Delta H_{\rm cloud}$,  changes with surface gravity
(Fig.~1 in Helling et al. 2010). 

The total number of potentially
released electrons per unit volume in a large-scale velocity field within each $\Delta h_{\rm i}$ is
then given by
\begin{equation}
\label{eq:ntotint}
n^{\rm tot}_{\rm col}=N\cdot \nu_{\rm col} \int_{\rm a_l}^{\infty} f(a) \frac{\Delta h_{\rm i}}{{\rm v}^{\rm sed}(a)}\,da.
\end{equation}
$N$ is the number of electrons produced per collision.  In case of
 a process of comparable efficiency such as secondary electron
emission, $N\sim 8$ elementary charges would be produced per collision
with a grain made of mixed materials (Sect.~\ref{ss:wf}). For other
processes, $N$ can be larger by orders of magnitude.   Also smaller $\Delta h_{\rm i}$ might reveal other, small-scale velocities as leading terms compared to the large-scale velocity considered in Eqs.~(\ref{eq:ntotint}) potentially increasing the number of collisions $n^{\rm tot}_{\rm col}$.

Using $f(a)= n_{\rm d}(a) \delta(a)$, Eq.~\ref{eq:ntotint} simplifies to
\begin{equation}
\label{eq:ntot}
n^{\rm tot}_{\rm col}=N\cdot \nu_{\rm col} n_{\rm d}(a) \frac{\Delta h_{\rm i}}{{\rm v}^{\rm sed}(a)}.
\end{equation}
The sedimentation velocity ${\rm v}^{\rm sed}(a)$ of the impinging
 (primary) grains is approximated by the equilibrium drift velocity in
 Eqs.~\ref{eq:vsed1} and~\ref{eq:vsed2}. 
 This approximation seems
 well justified, however, since the time for particles to accelerate
 to their terminal fall speed ${\rm v}^{\rm sed}(a)$ is small (Woitke
 \& Helling 2003). Even if grains are stopped by a collision, the
 grains will continue falling under the influence of gravity and they
 will achieve their terminal drift velocity again quickly.  This
 does not take turbulent mixing into account which would increase the sedimentation time. Large-scale
 convective replenishment is taken into account as part of the atmosphere simulation where  it impacts the dust formation process and
 by that the gravitational settling.

We assume that the dust and the gas (including thermal electrons) have
the same temperature, $T_{\rm gas}=T_{\rm dust}$.  Woitke \& Helling
(2003) showed that the liberation of latent heat of condensation and
the heating by friction is well balanced by radiative cooling and
inelastic collisions with the gas particles for grain sizes to be
expected in clouds of substellar atmospheres. We further assume that
the electrons thermalise with the gas or dust, hence $T_{\rm e}=T_{\rm
gas}$.  However, $T_{\rm e}=T_{\rm gas}$ may lead to an
underestimation of the electron pressure $p_{\rm e}$ as electrons tend
to have higher kinetic energies than $T_{\rm gas}$ in particular when
originating from non-equilibrium processes like in streamers.  For a
first approximation, we retain this assumption and calculate an
electron pressure equivalent,
\begin{equation}
\label{eq:petot}
p_{\rm e,col}=n_{\rm col}^{\rm tot}kT_{\rm gas} \quad .
\end{equation}

We have shown in Sect.~\ref{sss:en} that only turbulence-enhanced
dust-dust collisions are capable of achieving appropriate relative
grain-grain velocities to free electrons from the outer grain lattice
(Fig.~\ref{fig:col_energy}). These grains have sizes between $\approx
0.01\mu$m and $\approx 0.5\mu$m (Fig.~\ref{fig:grain_size}). They are made of a mix of silicates
(Mg$_2$SiO$_4$[s], MgSiO$_3$[s], SiO$_2$[s]) which takes up
$\approx$80\% of the grain volume with $\approx$15\% iron inclusions
(Figs.~\ref{fig:Vd5},~\ref{fig:Vd3}). 

\begin{figure}
\begin{center}
\resizebox{11cm}{!}{\includegraphics{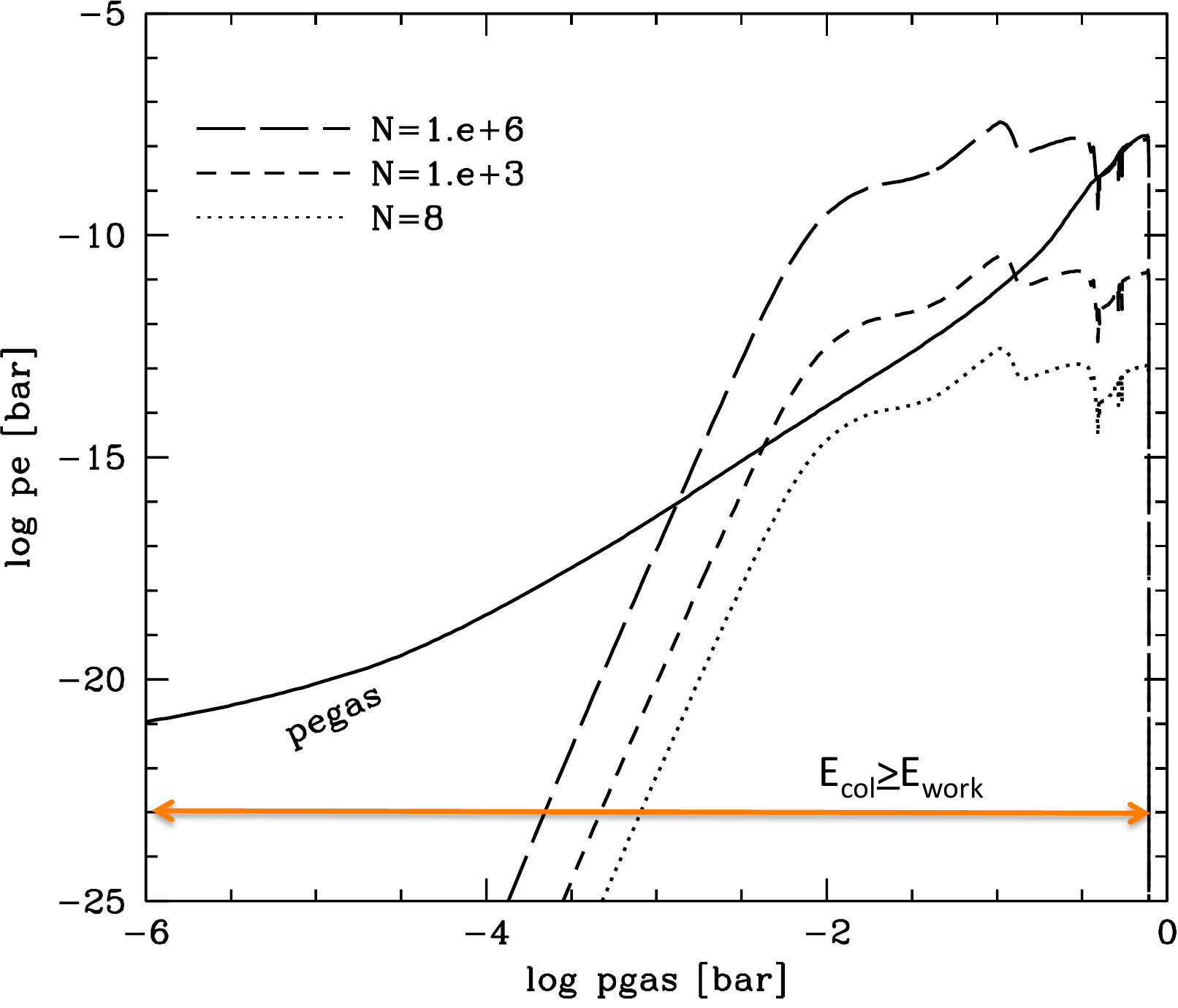}}\\
\resizebox{11cm}{!}{\includegraphics{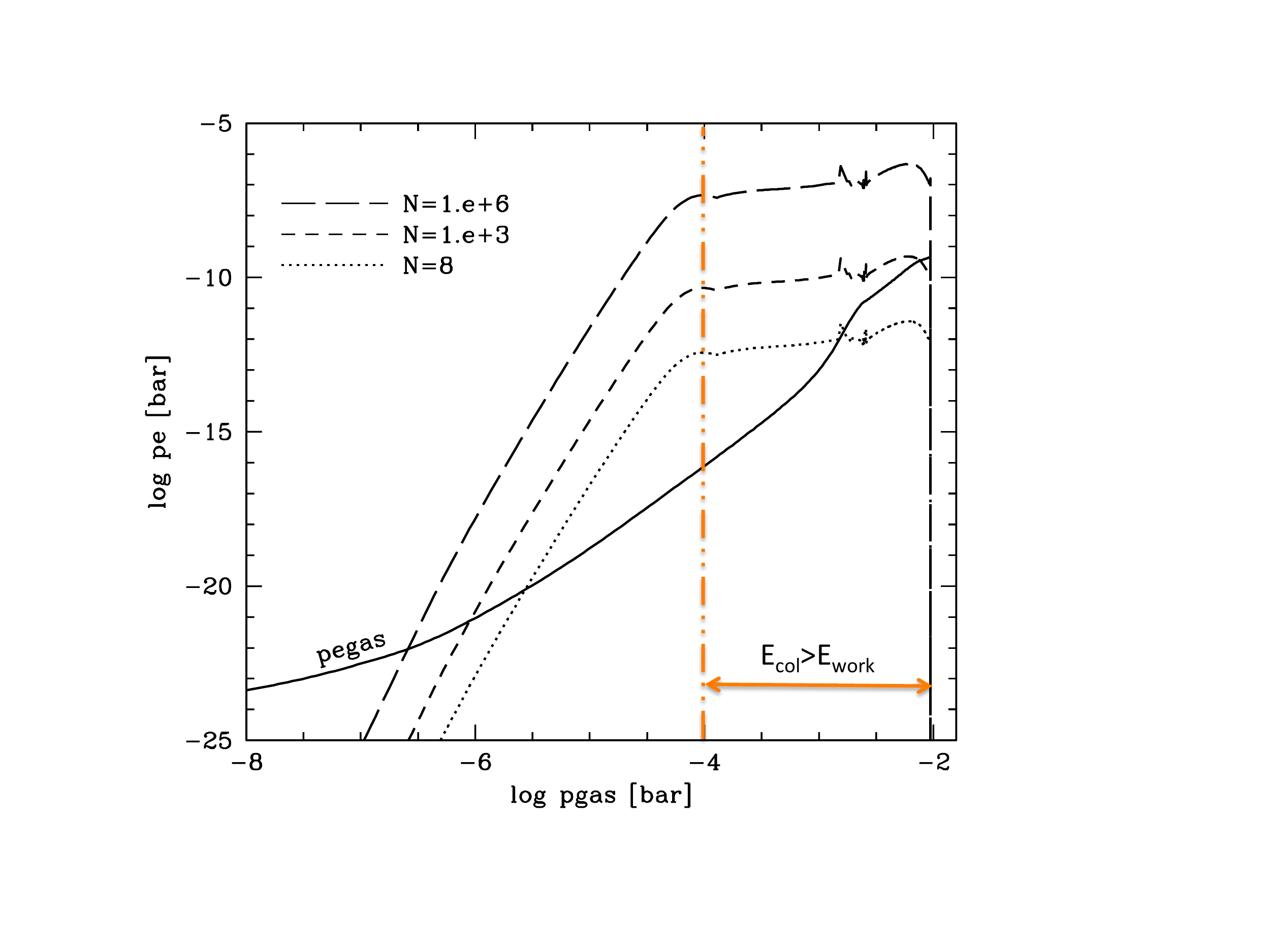}}
\caption{\small Electron pressure resulting from secondary electron
production of N=8 (dotted), N=$10^3$ (short dashed), and N=$10^6$
(long dashed) electrons per dust-dust collisions compared to the
atmospheric, thermal electron pressure (solid line) for two {\sc Drift-Phoenix} atmosphere (T$_{\rm
eff}=1600$K, log(g)=5.0 (top) / log(g)=3.0 (bottom), solar metallicity).  The horizontal arrow shows in which p$_{\rm gas}$-range  dust-dust collisions are energetic enough such that $E_{\rm col}>E_{\rm work}$.} \label{fig:p_e}
\end{center}
\end{figure}

Figure~\ref{fig:p_e} compares the atmospheric, thermal electron
pressure $p_{\rm e}$ as result of the {\sc Drift-Phoenix} atmosphere
simulation (solid black line) with the electron pressure $p_{\rm
e,col}$ resulting from turbulence-enhanced dust-dust collisions
producing N=8 charges (guided by the example of secondary electron production, dotted line),
N=$10^3$ (dashed line, Desch \& Cuzzi 2000), and N=$10^6$.  Hence,
 if the number of charges produced during the dust-dust collisions is comparable to the low efficiency of secondary electron production, 
the local
electron pressure will not be affected in a brown dwarf atmosphere. It, however, can
resupply the surface with charges, a precondition need for electron
avalanche process and streamers as suggested by Helling et
al. (2010).  Other mechanisms are needed to increase the number of
electrons released from the grain lattice to allow $p_{\rm e,col} \gg
p_{\rm e,gas}$. Helling et al. (2010) demonstrated that a release of
N=$10^6$ electrons per atmosphere layer would allow to couple 50\% of
the cloud volume to a large scale magnetic field.  We discussed
electron avalanche processes as attractive possibility to achieve such
a high number of free charges. In Sect.~\ref{ss:regimes}, we demonstrate that in which atmospheric regimes superposition of avalanche processes during dust-dust collisions are likely.

 In gas planet atmospheres, the situation becomes more favourable
for dust-dust collisions to increase the local degree of gas
ionisation as the thermal gas pressure is lower than in brown dwarf
atmospheres. Also the atmospheric volume affected is larger as the
cloud has a larger geometrical extension than in brown dwarfs.

 \section{Discussion}\label{ss:regimes}

\subsection{Lightning vs. coronal discharge}

\begin{figure}
\begin{center}
\resizebox{11cm}{!}{\includegraphics{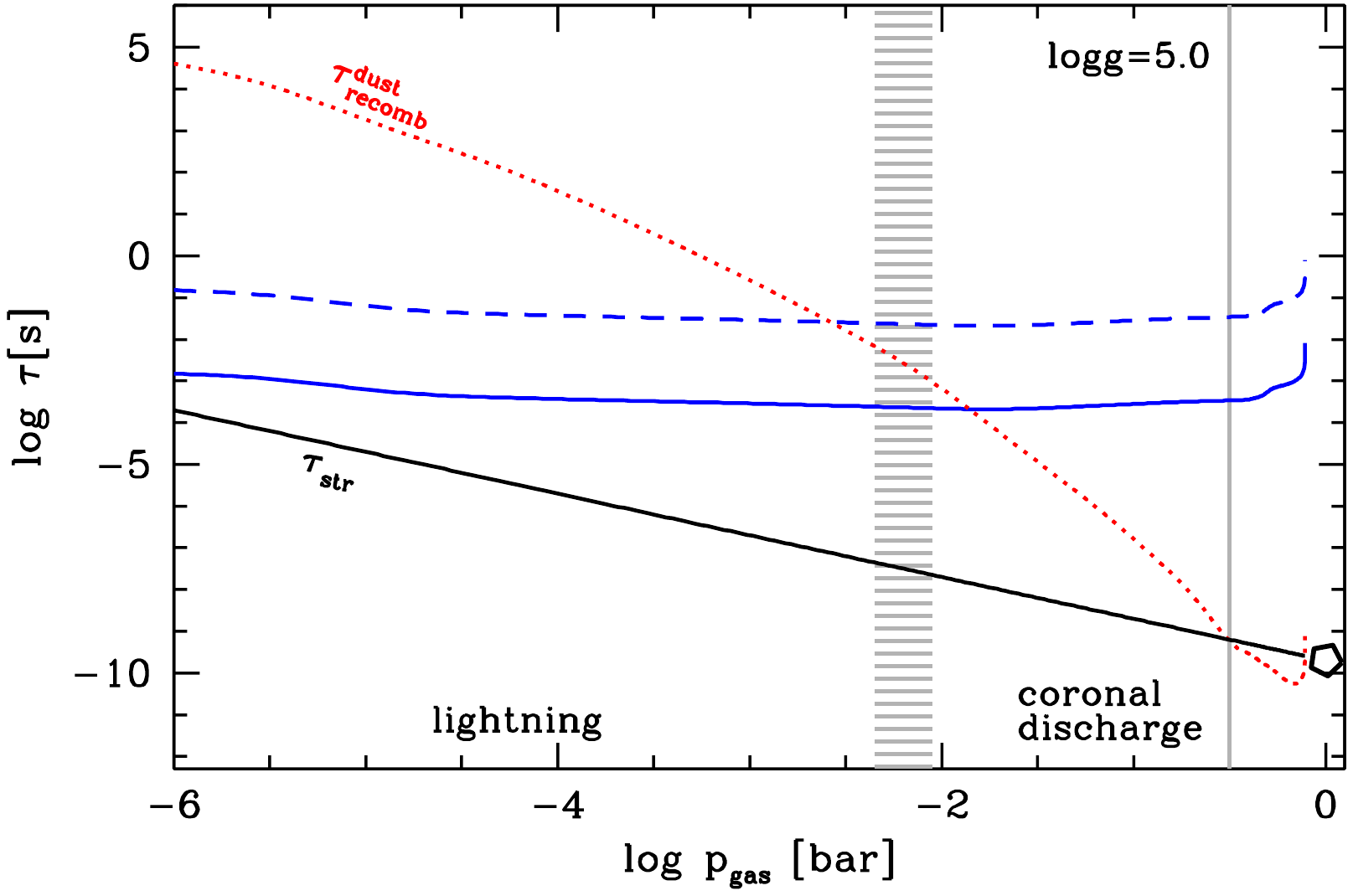}}\\
\resizebox{11cm}{!}{\includegraphics{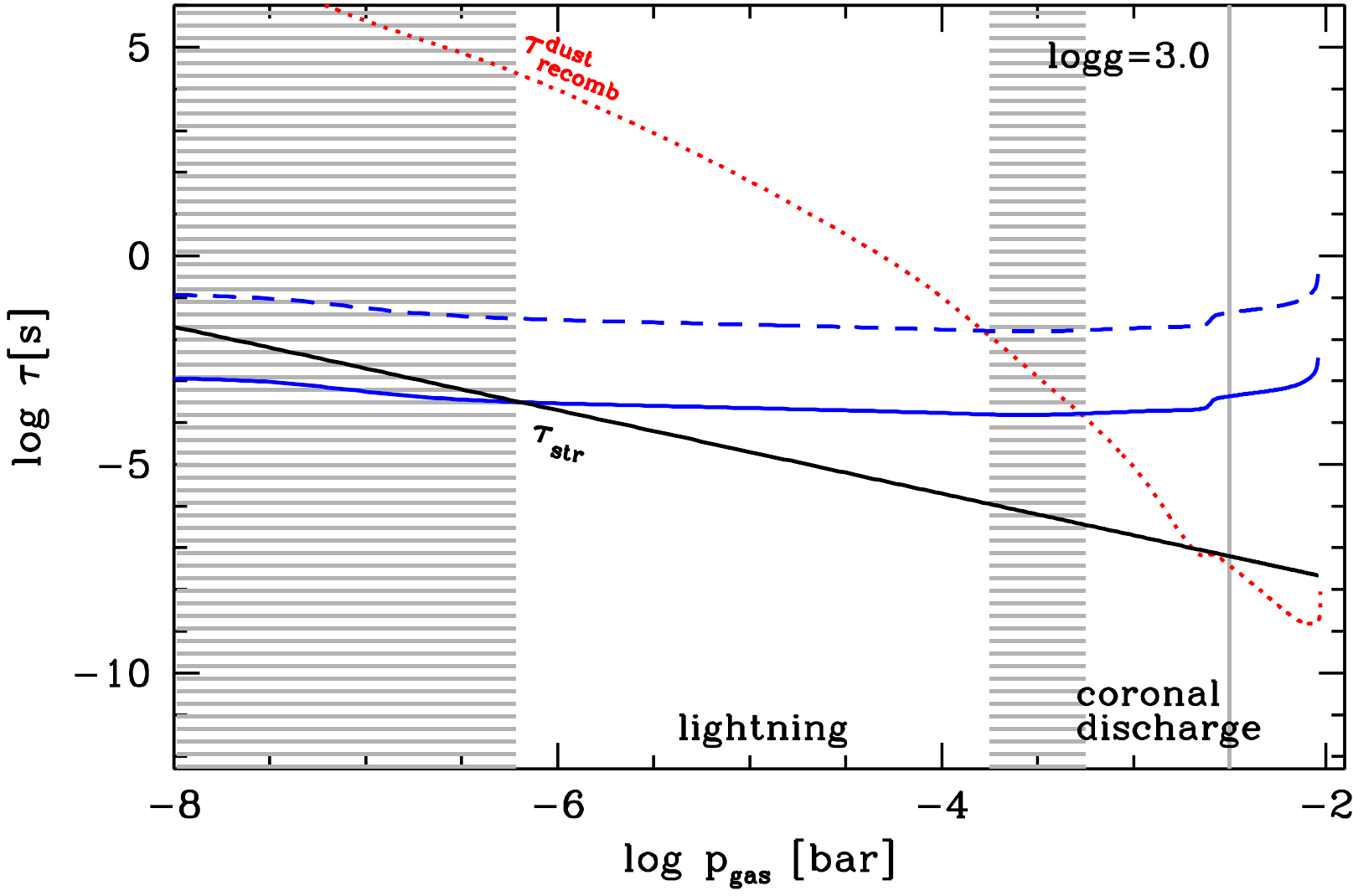}}
\caption{\small Time scale comparison of a streamer establishing
($\tau_{\rm str}$, black solid line), recombination onto the dust
grain ($\tau_{\rm recom}^{\rm dust}$ for $q=10$e, red dotted line;
Fig. 4 in Helling et al. 2010), and streamer superposition
time scale $t=n_{\rm d}^{-1/3} / v^{\rm sed}$ (blue solid, blue dashed
for $10^2 \times v^{\rm sed}$) based on the mean grain sizes from the {\sc
Drift-Phoenix} atmospheres (T$_{\rm eff}=1600$K, log(g)=5.0 /
log(g)=3.0, solar metallicity). Multiple encounters of streamer
electron clouds with the electric field of charged grains can happen in
the 'lightning'-regime, but not in the 'coronal discharge'-regime. The extension of the 'lightning'-regime decreases if turbulence slows down the grain settling process (blue solid line). }
\label{fig:taus}
\end{center}
\end{figure}

We have demonstrated so far that dust-dust collisions can be energetic
enough to release electrons from the cloud particle's lattice. These
electrons can be expelled into the gas phase or can remain on the
grain surface. In the case of the electron remaining on the grain's
surface, an electric field will build up which can be strong enough to
cause an avalanche process that produces an exponentially increasing
number of free electrons. As the recombination time back onto the
grain surface is rather large, these electrons can exist for a time in
the gas phase, and hence locally increase the degree of ionisation for
a certain time (Figs.~\ref{fig:p_e},~\ref{fig:taus}). But with which
efficiency does this occur in the cloud and could it lead to the
occurrence of lightning? A superposition of avalanche-streamer
processes will lead to more and more free electrons for a short time
period which then may be defined as lightning. An estimate of the time
scale, $t=n_{\rm d}^{-1/3} / v^{\rm sed}$, on which dust particles pass
through such a previously formed electron cloud, and hence,
potentially initiate another electron avalanche, shows that a certain
fraction of the cloud is prone to lightning-like discharge events in
brown dwarfs and in gas planets, namely where $\tau_{\rm
str}<t<\tau_{\rm recom}^{\rm dust}$ in Figs~\ref{fig:taus}. The
borders of this lightning region are not very clearly defined as
turbulence will decrease the effective sedimentation velocity
(blue solid vs. dashed lines).  If dust particles pass an electron
cloud slowly and the electrons can recombine, lightning is less likely
and a less powerful coronal discharge-like behaviour on smaller
scales should be expected as long as $\tau_{\rm str}<\tau_{\rm
recom}^{\rm dust}$.

The lightning and the coronal discharge regimes will exist in both brown dwarfs
and in gas planet atmospheres. Our results suggest a hierarchy of lightning  and coronal discharges with the lightning occurring at lower pressures  in the upper part of the cloud.  However, the maximum of the collisional
dust-dust energies (Fig.~\ref{fig:col_energy}) is more likely to be
located in the coronal discharge regime in our planetary atmosphere
example, while dust-dust collisions act across both regimes
in the brown dwarf example. How would then the cloud particles be
charged if dust-dust collisions are unfavourable inside a potential lightning
regime? Cosmic ray ionisation maybe a possibility.

Cosmic rays (CR) seem an important source of atmospheric ionisation in
the solar system planets. Observation of Earth clouds, however,
suggest that the actual charge production is not overly efficient but
nevertheless important for coagulation processes. Nicoll \& Harrison
(2010) determine $17\,\ldots\,150$e per particle in cloud edges on
Earth which can be directly related to ionisation by cosmic rays. The
charging of the cloud particles is, however, not a direct result of
the impact of the high energy CR particle on the cloud but rather of the ion
current that develops from the CR ionisation of the gas above the
cloud. These ions attach to the cloud particles. The question is
whether galactic cosmic rays could be a global source of ionisation
for extrasolar low-mass objects.  

\section{Conclusions}\label{sec:concl}
Dust clouds are an integral part of the atmospheres of very low mass
objects like Brown Dwarfs and planets.  Clouds determine the local
chemistry by element consumption and they influence the radiative and
convective energy transport by their large opacity in the
atmospheres. The aim of this paper is to demonstrate that dust grains
in Brown Dwarf atmospheres can be charged, and to investigate whether
the presence of dust in Brown Dwarf atmospheres can contribute to its
overall ionization level, a necessary condition for magnetic coupling
of the atmosphere. For this purpose, we focused on collisional
processes of the dust phase to cause additional ionization of the
atmosphere, an aspect that has not yet been considered in earlier
research.  We, however, acknowledge that a large variety of micro-physical  processes can be involved into the ionisation of a mineral cloud which have not yet been taken into account in our model. We find that collisional energies can be high enough to
ionize the dust phase over the whole extension of atmospheric clouds
of a late type Brown Dwarf (T$_{\rm eff}$=1600, $\log(g)=5$) and over a large part in a giant
planet's atmosphere (T$_{\rm eff}$=1600, $\log(g)=3$) if the influence of
the turbulent gas motion on the dust grains is taken into account. We
interpret our results such that collisional ionization of the dust
grains on its own does not provide, in the first place, an ionization
level which is sufficiently high for magnetic coupling.  A
collisionally charged dust phase could, however, trigger secondary non
linear charging processes such as electron avalanches that lead to
lightning: In the case that the charging process is rather a charge
exchange between colliding dust grains, the larger grains will be
charged positively and the smaller grains would carry away the
negative charges. This may lead to large scale charge separation due
to differential sedimentation and also differential response to the
turbulent gas drag of grains of different mass/size.

We suggest that although thermal gas ionisation may decline in objects
across the fully-convective boundary, dust ionisation may take over in
the lowest mass objects. The onset of atmospheric dust formation, the
cloud depth and its particle characteristics may therefore correlate
with the anomalous X-ray and radio emission in atmospheres that are
cool, but contain highly charged clouds.

In the context of this paper, it may be surprising that only a
fraction of Brown Dwarfs seem to show activity in form of X-ray or
radio emission and not nearly as much is known for extrasolar planets.
The observed intermittency of the X-rays might be interpreted as sign
for an intermittent dust cloud distribution as suggested by us in
earlier works (Helling et al. 2004).  Depending on the objects
parameter like mass, effective temperature, metallicity, age, the
atmosphere does change and so do the conditions for electrification
which we are only beginning to explore. Our first study does suggest
that cloud electrification is more pronounced in giant planets
compared to the more compact brown dwarf atmospheres. However, both
brown dwarfs and giant planets are prone to lightning events inside
their dust clouds.  Systematic investigation of this are
part of our future work.


{\bf Acknowledgement:} We thank the anonymous referee for a very constructive refereeing process. ChH highlights financial support of the European Community under the FP7 by an ERC starting grant, and the hospitality of the KITP Santa Barbara, University of California during the program 'Theory and Observation of Exoplanets'. FM
acknowledges support by the DFG cluster of excellence 'Origin and
Structure of the Universe'.

\label{lastpage}

\clearpage




\clearpage

\clearpage
\clearpage

\clearpage
\clearpage

\clearpage







\clearpage


\begin{thebibliography}{}
	
\bibitem{} Antonova A., Doyle J.G., Hallinan G., Bourke S., Golden A. 2008, A\&A 487, 317

\bibitem{} Baron E., Hauschildt P.~H., Allard F., et~al. 2003, in IAU
  Symposium, Vol. 210, Modelling of Stellar Atmospheres, 19
  
\bibitem{} Berger E., Basri G., Fleming T.A., Giampapa M.S., Gizis J.E. et al. 2010, ApJ 709, 332

\bibitem{} Berger E., Gizis J. E., Giampapa M. S., Rutledge R. E., Liebert J et al. 2008, ApJ 673, 1080

\bibitem{} Berdyugina S.V., Berdyugin A.V., Fluri D.M., Piirola V. 2011, ApJ 728, 6

\bibitem{} Burgasser A.J., Simcoe R.A., Bochanski J.J. et al. 2010, ApJ 725, 1405

\bibitem{} Burrows A., Sudarsky D., Hubeny I. 2006, ApJ 640, 1063

\bibitem{} Chabrier G., Baraffe I., Allard F., Hauschildt P. 2000, ApJ 542, 464

\bibitem{} Christensen U.R.,  Holzwarth V., Reiners A. 2009, Nature, 457, 167

\bibitem{} Currie Th., Burrows A.S., Itoh Y. et al. 2011, ApJ (arXiv1101.1973)

\bibitem{} Dehn M. 2007, PhD Thesis, University Hamburg

\bibitem{} Desch S.J., Cuzzi J.N. 2000, Icarus 143, 87

\bibitem{} Diver D.A., Clarke D. 1996, J.Phys. D: Appl. Phys. 29, 687

\bibitem{} Dowds B.J.P., Barrett R.K., Diver D.A. 2003, Physc Rev E 68, 026412


\bibitem{} Freytag A., Allard F., Ludwig H.-G., Homeier D., Steffen M. 2010, A\&A 513, 19

\bibitem{} Fortov V.E., Nefedov A.P., Molotkov V.I., Poustylnik M.Y., Torchinsky V.M. 2001, PhysRevLet 87, 
205002-1


\bibitem{} G\"uttler C., Blum J., Zsom A., Ormel C.W., Dullemond C.P. 2010, A\&A 513, 56

\bibitem{} Hallinan G., Antonova A., Doyle J.G., Bourke S., Lane C., Golden A. 2008, ApJ 684, 644

\bibitem{} Hallinan G., Antonova A., Doyle J.G., Bourke S., Brisken W.F., Golden A. 2006, ApJ 653, 690

\bibitem{HB99} {Hauschildt}, P.~H. \& {Baron}, E. 1999, Journal of Computational and Applied
  Mathe\bibitem{} matics, 109, 41

\bibitem[2010] {} Helling, Ch., Jardine M., Witte S., Diver D.A. 2010, ApJ, in press

\bibitem[2009] {} Helling Ch., Rietmeijer F.J.M 2009, IJAsB 8, 3


\bibitem[2008] {} Helling Ch., Ackerman A., Allard F., Dehn M., Hauschildt P., Homeier D., Lodders K., Marley M. et al. 2008, MNRAS 391, 1854

\bibitem[2008] {} Helling Ch., Woitke P., Thi W.-F. 2008, A\&A 485, 547

\bibitem[2007] {} Helling Ch., Dehn M., Woitke P., Hauschildt P.H. 2008a, ApJ 675, L105
\bibitem[2007] {} Helling Ch., Dehn M., Woitke P., Hauschildt P.H. 2008b, ApJ 677, L157

\bibitem{} Helling Ch. 2005, in "Interdisciplinary Aspects of Turbulence", Kupka F., Hillebrandt W. (eds), 152

\bibitem{} Helling Ch., Klein R., Woitke P., Nowak U., Sedlmayr E. 2004, A\&A 423, 657
\bibitem{} Helling Ch., Oevermann M., L\"uttke M.J.H., Klein R., Sedlmayr E. 2001, A\&A 376, 194


\bibitem{} 	Kirkpatrick J. D. 2005, ARA\&A 43, 195

\bibitem{} Kopnin S.I., Kosarev I.N., Popel S.I., Yu, M.Y. 2004, Planetary and Space Science 52, 1187

\bibitem{} Li E., Ebert U., Brok W.J.M. 2008, IEEE Trans. Plasma Sci., 36, 914


\bibitem{Morfill85.1}
G.~E. {Morfill}.
\newblock {Physics and chemistry in the primitive solar nebula.}
\newblock In {\em Birth and infancy of stars}, pages 693--792, 1985.


\bibitem{} Nicoll K.A., Harrison R.G. 2010, GRL 37, L31802

\bibitem{} Niedrig H., 1992, Spinger text book, Berlin

\bibitem{} Pinto C., Galli D. 2008, A\&A 484, 17

\bibitem{} Pont F., Knutson H., Gilliland R. L., Moutou C., Charbonneau D. 2008, MNRAS 385, 109

\bibitem{} Poppe T., Schr\"apler R. 2005, A\&A 438, 1

\bibitem{} Reiners A., Basri G.2008, ApJ 684, 1390

\bibitem{}  Rosenberger 2001; Astrophysics and Space Science 277, 125

\bibitem{} Saumon D., Marley M. S. 2008, ApJ 689, 1327

\bibitem{}  Scholz \& Eisl\"offel, 2004, A\&A, 421,259-271

\bibitem{} Sinclair J.A., Helling Ch., Greaves J.S. 2010, MNRAS 409, 49

\bibitem{}  Sickafoose A. A., Colwell J. E., Horanyi M., Robertson S. 200, Phys Rev Lett 84, 6034

\bibitem{} Stark C.R., Potts H.E., Diver D.A. 2006, A\&A 457, 365

\bibitem{} Tsuji T. 2002, ApJ 575, 264

\bibitem{} Umebayashi T. and Nakano T. 2009, ApJ 690, 69

\bibitem{} V{\"o}lk H.J., Jones F.C., Morfill G.E., Roeser S. 1980, A\& A 85, 316

\bibitem{} Witte S., Helling Ch., Hauschildt P.H. 2009 A\&A 506, 1367
\bibitem{} Witte S., Helling Ch., Hauschildt P.H. 2011, in press

\bibitem{Woitke03.1}
{Woitke} P., {Helling} Ch
\newblock {A\& A}, 399:297--313, February 2003.



\end{thebibliography}
\end{document}